\documentclass{aa}  
\usepackage{graphicx}
\usepackage{scalefnt}
\usepackage{lscape}
\usepackage{multirow}
\usepackage{subfig}
\usepackage[multiple]{footmisc}
\usepackage{txfonts}
\usepackage{aalongtable}

\begin{document}

%
\newbox\grsign \setbox\grsign=\hbox{$>$} \newdimen\grdimen \grdimen=\ht\grsign
\newbox\simlessbox \newbox\simgreatbox
\setbox\simgreatbox=\hbox{\raise.5ex\hbox{$>$}\llap
     {\lower.5ex\hbox{$\sim$}}}\ht1=\grdimen\dp1=0pt
\setbox\simlessbox=\hbox{\raise.5ex\hbox{$<$}\llap
     {\lower.5ex\hbox{$\sim$}}}\ht2=\grdimen\dp2=0pt
\def\simgreat{\mathrel{\copy\simgreatbox}}
\def\simless{\mathrel{\copy\simlessbox}}
\newbox\simppropto
\setbox\simppropto=\hbox{\raise.5ex\hbox{$\sim$}\llap
     {\lower.5ex\hbox{$\propto$}}}\ht2=\grdimen\dp2=0pt
\def\simpropto{\mathrel{\copy\simppropto}}
\makeatletter
\makeatother 
\title{High-resolution abundance analysis of four red giants in  the globular cluster NGC 6558
\thanks{Observations collected at the European Southern Observatory,
Paranal, Chile (ESO), under programme 93.D-0123A.}}
\author{
B. Barbuy\inst{1}
\and
L. Muniz\inst{1}
\and
S. Ortolani\inst{2,3}
\and
H. Ernandes\inst{1}
\and
B. Dias\inst{4,5}
\and
I. Saviane\inst{4}
\and
L. Kerber\inst{6}
\and
E. Bica\inst{7}
\and
A. P\'erez-Villegas\inst{1}
\and
L. Rossi\inst{8}
\and
E.V. Held\inst{9}
}
\offprints{B. Barbuy}

\institute{
Universidade de S\~ao Paulo, IAG, Rua do Mat\~ao 1226,
Cidade Universit\'aria, S\~ao Paulo 05508-900, Brazil;
e-mail: barbuy@astro.iag.usp.br
\and
Dipartimento di Fisica e Astronomia, Universit\`a di Padova, I-35122 Padova,
 Italy
\and
INAF-Osservatorio Astronomico di Padova, Vicolo dell'Osservatorio 5,
I-35122 Padova, Italy
\and
European Southern Observatory, Alonso de Cordova 3107, Santiago, Chile
\and
Facultad de Ciencias Exactas, Universidad Andr\'es Bello, 
Av. Fernandez Concha 700, Las Condes, Santiago, Chile
\and
Universidade Estadual de Santa Cruz, DCET, Rodovia Jorge Amado km 16, 
Ilh\'eus 45662-000, Bahia, Brazil
\and
Universidade Federal do Rio Grande do Sul, Departamento de Astronomia,
CP 15051, Porto Alegre 91501-970, Brazil
\and
Centre for Astrophysics and Supercomputing, Swinburne University of Technology, Hawthorn, Victoria 3122, Australia
\and
INAF - Osservatorio Astronomico di Padova, vicolo dell'Osservatorio 5, 35122 Padova, Italy
}
 
   \date{Received; accepted }

\abstract
   {NGC 6558 is a bulge globular cluster with a blue horizontal branch (BHB), 
combined with a metallicity of 
[Fe/H]$\approx$-1.0. It is similar to HP~1 and NGC 6522, which could be among 
the oldest objects in the Galaxy. Element abundances in these clusters
 could reveal the nature of the first supernovae. }
   {We aim to carry out detailed spectroscopic analysis for four red giants of NGC 6558, 
 in order to derive the abundances of the light elements
C, N, O, Na, Al, the $\alpha$-elements Mg, Si, Ca, Ti, and the
 heavy elements Y, Ba, and Eu.}
   {High-resolution spectra of four stars  with FLAMES-UVES@VLT UT2-Kueyen were analysed. 
  Spectroscopic parameter-derivation was based on excitation
 and ionization equilibrium of \ion{Fe}{I} and \ion{Fe}{II}.}
   {This analysis results in a metallicity of  [Fe/H] = $-1.17\pm0.10$ for NGC 6558. 
We find the expected $\alpha$-element enhancements in O and Mg with
 [O/Fe]=+0.40, [Mg/Fe]=+0.33,
and low enhancements in Si and Ca. Ti has a moderate enhancement of 
[Ti/Fe]=+0.22.
The r-element Eu 
 appears very enhanced with a mean value of [Eu/Fe]=+0.63.
Ba appears to have a solar abundance ratio relative to Fe.}
   {NGC~6558 shows an abundance pattern that could be typical of
the oldest inner bulge globular clusters, together with
the pattern in the similar clusters NGC~6522 and HP~1.
They show low abundances of the
 odd-Z elements Na and Al, and of the explosive nucleosynthesis
$\alpha$-elements Si, Ca, and Ti. The hydrostatic burning $\alpha$-elements
O and Mg are normally enhanced as expected in old stars enriched with yields
from core-collapse supernovae, and 
 the iron-peak elements Mn, Cu, Zn show low abundances, which is expected
for Mn and Cu, but not for Zn. Finally, the cluster trio
NGC 6558, NGC 6522, and HP~1 show similar abundance patterns. }
\titlerunning{Abundance analysis of  red giants in NGC 6558}
\authorrunning{B. Barbuy et al.}
\maketitle
%
\section{Introduction}

Globular clusters in the Galactic bulge are probes of the formation
processes of the central parts of the Galaxy.
The first dwarf-like galaxies were likely  the formation-sites of
the earliest globular clusters, and they were subsequently incorporated
in the halo and bulge of galaxies such as the Milky Way 
(Nakasato \& Nomoto 2003; Boley et al. 2009; Bromm \& Yoshida 2011; 
Tumlinson 2010). Therefore, metal-poor inner bulge globular clusters might be 
relics of an
early generation of long-lived stars formed in the proto-Galaxy. 

Barbuy et al. (2018a) report a concentration
and even dominance of clusters with [Fe/H]$\sim$-1.00 within
about 10$^{\circ}$ of the Galactic centre.
A metallicity of -1.3$\simless$ [Fe/H] $\simless$-1.0 is
 compatible with the lowest metallicities  of the bulk
of stellar populations in the Galactic bulge,
as shown by the metallicity distribution of bulge stars by
Zoccali et al. (2008, 2017), Hill et al. (2011),
 Ness et al. (2013), and Rojas-Arriagada et al. (2014, 2017).
This is due to a fast chemical enrichment 
in the Galactic bulge, as discussed in Chiappini et al. (2011),
Wise et al. (2012) and Cescutti et al. (2018).
 Very low fractions of more metal-poor stars were found so far
(e.g. Garc\'{\i}a-Perez et al. 2013, Casey \& Schlaufman 2015,
 Howes et al. 2016, Koch et al. 2016) --  see also the
discussion in Barbuy et al. (2018a).

Therefore it is crucial to  study such moderately metal-poor clusters
([Fe/H]$\sim$$-$1.0) located in the Galactic bulge, 
in order to further identify their signatures, which correspond
to the earliest stages of the Galactic bulge formation.
Their chemical abundances can reveal the nature of the first massive
stars and supernovae.

One candidate Milky Way relic is NGC 6558, which
 has a blue horizontal branch (BHB),
together with a metallicity of [Fe/H]$\sim$-1.0 (Rich et al. 1998; 
Barbuy et al. 2007),
with characteristics very similar to NGC~6522 
(Barbuy et al. 2009, 2014, Ness et al. 2014),
 and HP~1 (Barbuy et al. 2006, 2016). 
The trio NGC 6558, NGC 6522, and HP 1 are the main representatives
of such clusters in the inner 6$^{\circ}$ of the Galactic bulge, 
as can be seen in the classification given in Barbuy et al. (2009) --
see also Bica et al. (2016).
Their combined features indicate a very old age. 
The ages of NGC~6522 and HP~1 were recently confirmed to be 
very old, calculated to be around 13 Gyr by
Ortolani et al. (2011) and Kerber et al. (2018a,b). 

The cluster NGC 6558 is projected on the Galactic bulge, 
in a field described as an extended clear region by Blanco (1988).
The equatorial coordinates are (J2000) 
$\alpha$ = 18$^{\rm h}$ 10$^{\rm m}$18.4$^{\rm s}$, 
$\delta$ = -31$^{\circ}$ 45' 49", and
the Galactic coordinates are l = 0.201$^{\circ}$, b = -6.025$^{\circ}$.
NGC 6558 is very concentrated, with post-core collapse structure, a core radius
of r$_{\rm c}$ = 2'' and a half light radius r$_{\rm h}$ = 129'' according to
 Trager et al. (1995).

In Rossi et al. (2015), NGC 6558 was studied with a set of 
subarcsec ($\sim$0.5'') seeing images obtained at the
New Technology Telescope (NTT),  with a time 
difference of 19 years. This has allowed us to apply proper motion
decontamination, with high accuracy, and a reliable selection of member stars. 
Using the proper motion determinations by Rossi et al. (2015), 
combined with the radial velocity from high-resolution spectroscopy 
 by Barbuy et al. (2007) and heliocentric distance from 
literature (Bica et al. 2006),  P\'erez-Villegas et al. (2018) carried out 
an orbital analysis of NGC 6558 in a bulge/bar Galactic potential 
and with a Monte Carlo method, taking into account
 the uncertainties in the observational parameters. 
The cluster shows a pro-grade orbit, with peri-galactic and apo-galactic
 distances of $\sim$0.13 and $\sim$2.50 kpc, respectively, and a  maximum 
height of  $<|z|_\mathrm{max}>$$\sim$1.4 kpc,
pointing out that the cluster is confined in the innermost 
Galactic region with a bar-shape in the x-y projection, and 
boxy-shape in the x-z projection (in a frame co-rotating with the bar), 
indicating that NGC 6558 is trapped by the Galactic bar.

In Barbuy et al. (2007) we carried out a detailed abundance analysis of five
stars with spectra obtained with the GIRAFFE spectrograph at the
Very Large Telescope (VLT),  from the ESO projects
 71.B-0617A, 73.B0074A (PI: A. Renzini)
(see Zoccali et al. 2008). 
A mean metallicity of  [Fe/H]=$-$0.97$\pm$0.15 was obtained for NGC 6558.
From CaT triplet lines based on FORS2@VLT spectra,
for a sample of 28 globular clusters, Saviane et al. 
(2012) derived [Fe/H]=$-$1.03$\pm$0.14 for NGC 6558.
Based on FORS2@VLT spectra in the optical, Dias et al. (2015, 2016) derived metallicities for
51 globular clusters, including the same stars in the same clusters as in Saviane et al. (2012).
A method of full spectrum fitting from observations in the 
4600-5600 {\rm \AA} region was carried out. Recalling that the results depend
crucially on the library of spectra adopted, the final metallicity
of  [Fe/H]=$-$1.01$\pm$0.10 for NGC 6558 was obtained
 essentially with the use of a library of
synthetic spectra (Coelho et al. 2005). 

In this work we present results from spectra obtained for four stars with
the FLAMES-UVES spectrograph at the VLT,  at a resolution R$\sim$45,000.
 A detailed abundance analysis  of the sample stars 
was carried out using MARCS model atmospheres (Gustafsson et al. 2008).

   Photometric and spectroscopic data are described respectively  in Sects. 2
and 3.  Photometric stellar parameters effective temperature and  gravity  
are   derived in  Sect. 4.  Spectroscopic parameters are
derived in  Sect.  5 and   abundance ratios are   computed
 in  Sect. 6. Results are discussed in Sect. 7 and
 conclusions are drawn in Sect. 8.

   \begin{table}
\caption{Log of photometric observations, carried out on 20-21 May 2012
at the NTT. }             
\label{logbook}      
\scalefont{1.0}
\centering                          
\begin{tabular}{cccccc}        
\hline\hline                 
\noalign{\smallskip}
\hbox{Filter} &\hbox{Exp. (s)} &\hbox{Seeing (``)} &\hbox{Airmass}& \\
\noalign{\smallskip}
\hline
\noalign{\smallskip}
V & 20 & 0.55 & 1.01 & \\
I & 10 & 0.50 & 1.01 & \\
V & 30 x 4  &  0.60 & 1.01 & \\
V & 300  &  0.60 & 1.01 & \\
I & 180 x 4  &  0.60 & 1.01 & \\
I & 20 x 4  &  0.60 & 1.01 & \\
V & 90  &  0.60 & 1.01 & \\
I & 60  &  0.60 & 1.01 & \\
\noalign{\smallskip}
\hline                                   
\end{tabular}
\end{table}

\section{Photometric observations} 
\label{photometry}

Colour-magnitude diagrams (CMDs) for NGC 6558 were presented by Rich et al.
 (1998) in V vs. V-I,
and in the near-infrared (NIR) by Davidge et al. (2004) in  K vs. J-K diagrams.
 Davidge et al. (2004)
estimated a metallicity of [Fe/H]=-1.5$\pm$0.5, from Flamingos-I spectra of
resolution of R=350, obtained with the Gemini-South telescope.
We carried out a membership check for their brightest stars, 
and found that their
star \#5 should be a member, located at the very tip of the red giant branch
at V-I = 2.55; star \#11 is saturated in our images, and could not be 
verified for membership.
More recently, NGC 6558 has been further studied in the near-infrared by 
Chun et al. (2010),
and in the optical by Alonso-Garc\'{\i}a et al. (2012). 

In Rossi et al. (2015) a proper-motion cleaned CMD of NGC 6558 was produced,
where the
first epoch observations of NGC 6558 were taken in June 15, 1993,
and the second epoch data in May 20, 2012.
Therefore a time baseline of 19 years is available for proper
motion decontamination.

In the present work we give further detail on the NTT data and reductions,
beyond those given in Rossi et al. (2015), where data and reductions were only briefly described and reported
in their Table 1.
For the first epoch, we used NTT at 
the European Southern Observatory - ESO (La Silla) 
equipped with EMMI in the focal reducer mode.
The red arm and the 2024x2024 pixels Loral
UV-coated CCD ESO \# 34 detector were employed. 
The pixel size is 15 $\mu$m, corresponding 
to 0.35'',  with a full field of 11.8'x11.8'. 
The log of these observations can be found in Rich et al. (1998).
 
For the second epoch,  we used NTT equipped with EFOSC2,
with CCD ESO \# 40, UV-flooded, MPP, and
2048x2048 pixels, with pixel size of  15 $\mu$m,
corresponding to 0.12''/pixel, binned in two by two.
The full field is 4.1'x4.1'.
The log of NTT 2012 observations is given in Table 1.

We have measured 20 Landolt stars (Landolt 1983, 1992) 
during the two photometric nights
used to define the calibration transformations, and some of them have
been observed repeatedly.
Daophot II was used to extract the instrumental magnitudes.
These magnitudes have been calibrated using
Landolt stars, resulting in: 

\par V = v + 0.04(V-I) + 28.51 $\pm 0.015 \rm mag$

\par I = i - 0.01(V-I) + 27.96 $\pm 0.015 \rm mag$

\noindent for exposure times of 15 sec and airmasses of 1.15.
Due to crowding effects the transfer of the aperture magnitudes from
standards to the field stars gives an additional  $\pm$0.03 mag, and
the final magnitude zero point uncertainty amounts to $\pm$0.04.
The atmospheric extinction was corrected with the
standard La Silla coefficients for EFOSC2 (C$_V$ = 0.16,
C$_I$ = 0.08 mag/airmass).

The present proper motion measurements were based on the two
best seeing second epoch sets of images. We used the remaining images
for a test on photometric error effects.

\begin{figure}
\centering
\includegraphics[scale=0.45]{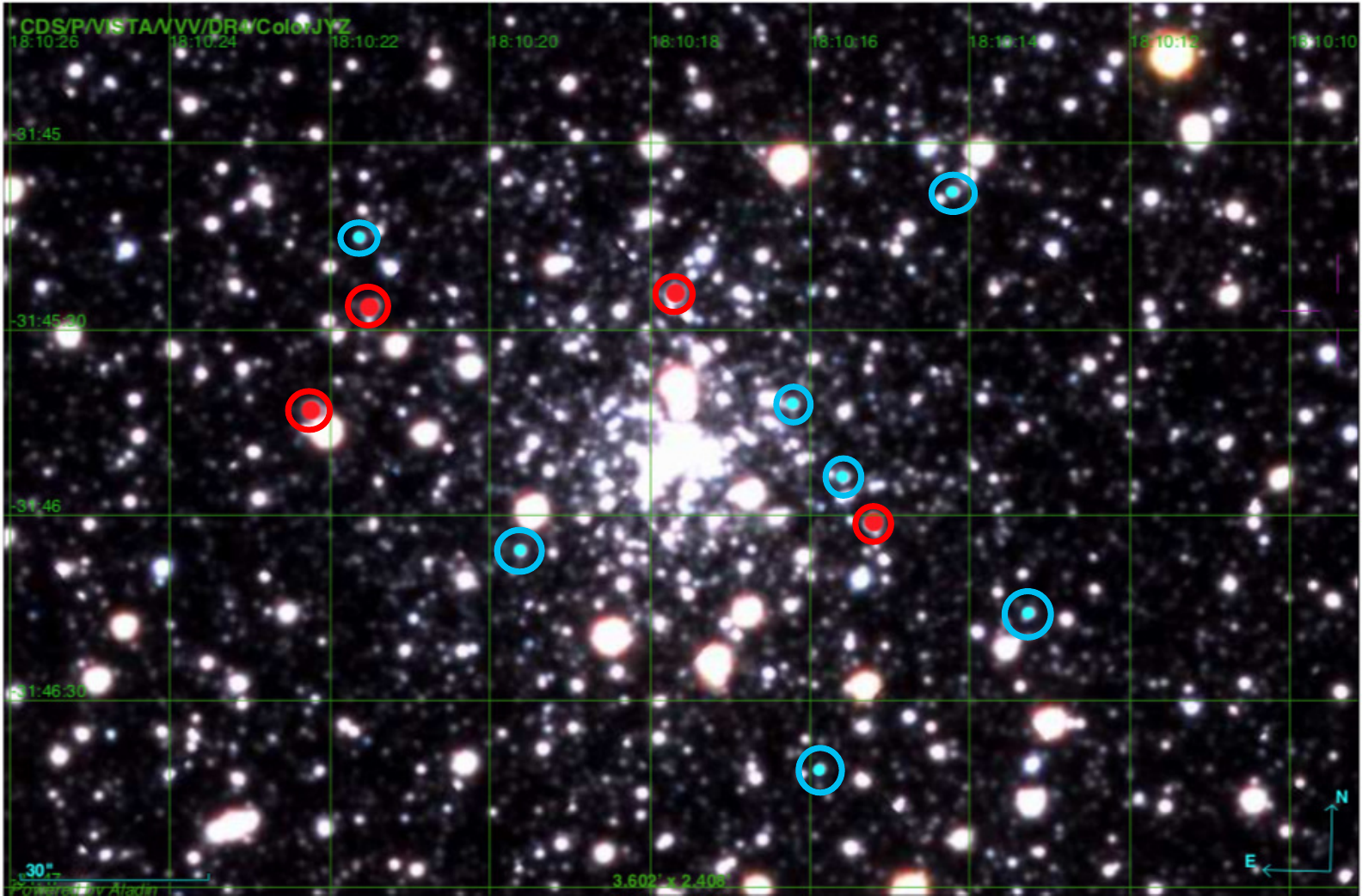}
\caption{VVV image of NGC 6558 in combined JYZ colours, in size of 3.6'x2.4'.
The four sample stars are identified as red filled circles, 
and RR Lyrae from OGLE
are identified by cyan filled circles. 
}
\label{n6558image} 
\end{figure}

In Figure \ref{n6558image} we show an image of NGC 6558 
combining J, Y, Z filters, from the Vista Variables in the Via Lactea
 (VVV\footnote{horus.roe.ac.uk/vsa/}; Saito et al. 2012) survey;
the four sample stars, and
the more central RR Lyrae stars identified in the 
Optical Gravitational Lensing Experiment 
(Udalski et al. 2002; Soszynski et al. 2014)
OGLE\footnote{ogledb.astrouw.edu.pl/$\sim$ogle/OCVS} 
are overplotted.

\begin{figure}
\centering
\includegraphics[scale=0.45]{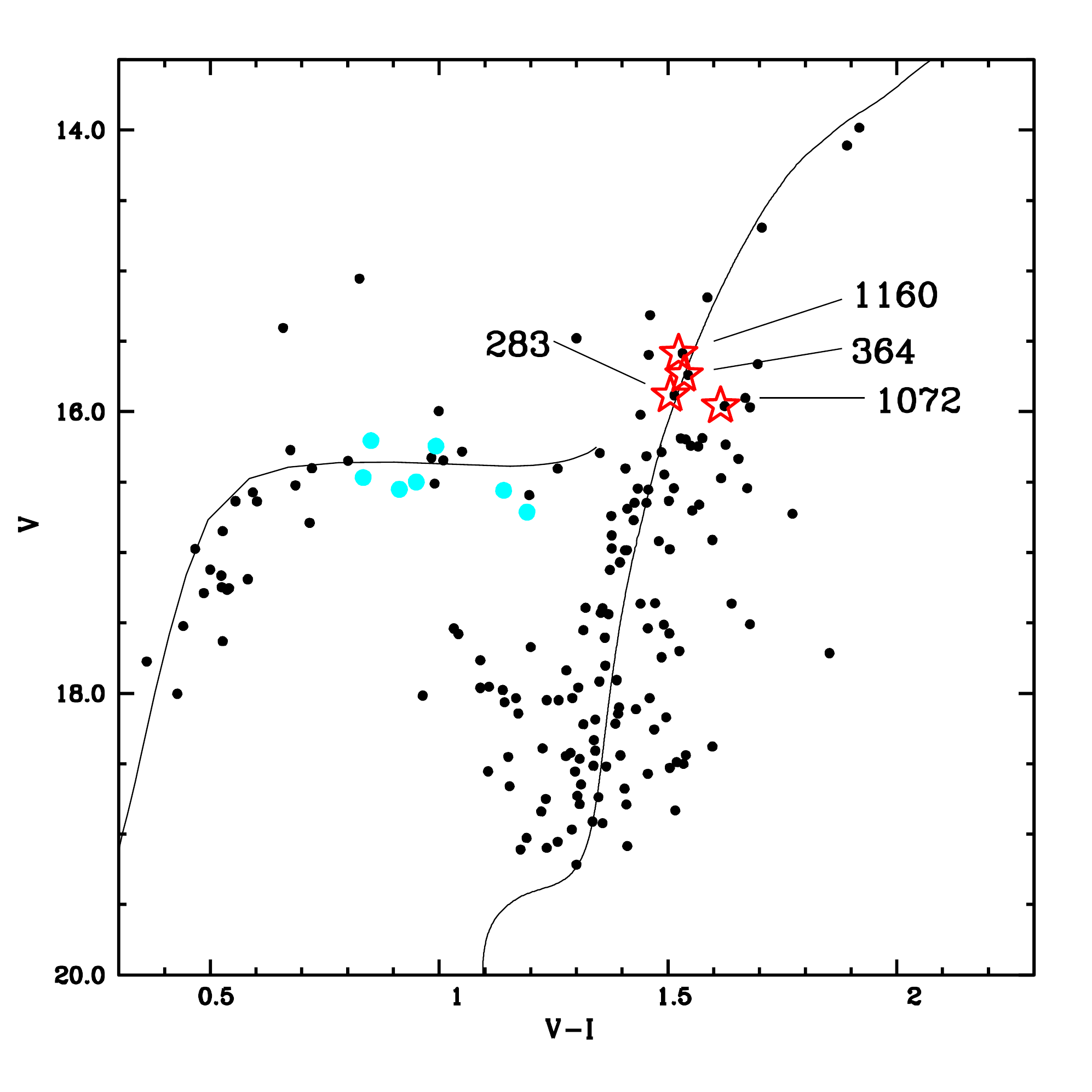}
\caption{$V$ vs. $V-I$ Colour-magnitude diagram of NGC 6558 from NTT 2012 data.
The four sample stars are shown as red stars, and RR Lyrae are shown as green filled 
circles. A BaSTI isochrone for an age of 13 Gyr,
[Fe/H]=-1.0 and primordial helium abundance (Y=0.25) is overplotted.
}
\label{cmdsergio} 
\end{figure}

In Figure \ref{cmdsergio} we give a  V vs. V-I proper-motion-cleaned CMD
(calibrated to the NTT 2012 data),
together with the BaSTI $\alpha$-enhanced 13 Gyr, [Fe/H]=-1.0,
and primordial helium (Y=0.25) isochrone fit
(Pietrinferni et al. 2004). The four sample stars and
the seven more central RR Lyrae from OGLE
are identified.
As can be seen, the four sample stars are red giant branch (RGB) stars,
and the RR Lyrae populate the instability strip of the horizontal
branch (HB).
 The BaSTI isochrones
employed are suitable to the subgiant and red giant branch
(SGB-RGB), and a separate isochrone
suitable to the zero-age horizontal branch (ZAHB) is included. 

\subsection{Reddening and distance}

In Table \ref{distance}, we report values of V magnitude of the
Horizontal Branch $V_{HB}$, reddening, distance modulus and distances
 adopted in the literature 
(Terndrup 1988, Hazen 1996, Rich et al. 1998, Davidge et al. 2004,
Barbuy et al. 2007,  Rossi et al. 2015), together with
the presently derived values. In a pioneer work Terndrup (1988) 
derived reddening and proper motion values for fields along
the bulge minor 
axis, including the NGC 6558 location at -6$^{\circ}$.
Distances of d$_{\odot}$ = 6.6 kpc and 6.3 kpc were found respectively by 
 Hazen (1996)
and Rich et al. (1998).   A revised distance taking into account
an updated absolute magnitude of the HB versus
metallicity relation by  Harris (1996,  updated in 2010)\footnote{
 http://www.physics.mcmaster.ca/Globular.html}
yields E(B-V)=0.44 and a distance to the Sun of d$_{\odot}$=7.4 kpc.
Davidge et al. (2004) adopt E(B-V)=0.44.
Barbuy et al. (2007) use photometric data from the 2.2m MPI telescope at ESO,
and obtain a magnitude of the HB level at  V$_{\rm HB}$ = 15.5.

 Alonso-Garc\'{\i}a et al. (2012) analyse the differential reddening
in NGC~6558, relative to a ridge line (not the absolute values).
For a mean reddening value the authors adopt a value from Schlegel 
et al. (1998), that is overestimated in particular in the
central regions,  because it measures the
 dust integrated in the line of sight all along the extent of the Galaxy.
Therefore these authors' mean value of E(B-V)=0.51 appears to be too high -
see literature values in Table \ref{distance}.
 Alonso-Garc\'{\i}a et al. find that  that NGC 6558 is a difficult
case (their Sect. 4.18), and they could not get a significant de-reddening improvement,
due to  the cluster being very contaminated. Their method is
also affected by stars with different reddening along the line of sight.

Rossi et al. (2015) use the 1993 NTT data combined with the 2012 NTT data,
and from these data a HB level at V$_{\rm HB}$ = 16.0, a
distance modulus of  ($m-$M)$_0$ = 14.43, A$_{\rm V}$ = 1.178,
and a distance of  d$_{\odot}$ = 7.4 kpc was obtained.
Conversion factors for E(V-I)/E(B-V) 
range from   1.31 (Dean et al. 1978),  
1.38 (Schlegel et al. 1998) to  1.40 (Schlafly \& Finkbeiner 2011). 
Schlegel et al. (1998) gives E(V-I)=0.53
towards NGC 6558, this being an upper limit, 
since it is measured all along the line of sight. 
OGLE reddening maps give E(V-I)=0.54 (Nataf et al. 2013),
Rich et al. (1998) derived E(V-I)=0.60, and
Harris (1996, updated in 2010) reports E(B-V)=0.44.
In the present work we
identified the HB at V(HB)=16.37, and find E(V-I)=0.50 (Table 2), 
which together with
reddening law values of E(V-I)/E(B-V)=1.33, and a total-to-selective  
absorption R$_{\rm V}$=3.1, give a V-band extinction of A$_{\rm V}$=1.17, 
and a distance  of d$_{\odot}$ = 8.26 kpc (Table \ref{distance}),
 farther than in previous measurements.

An additional constraint to the cluster distance can be obtained
 from the RR Lyrae. According to the M$_{\rm V}$-[Fe/H] relation derived 
by the Gaia collaboration (2017), RR Lyrae stars with [Fe/H]=-1.17 
(Sect. 5, Table \ref{tabteff}),
have M$_{\rm V}$$\sim$0.65$\pm$0.10.
Assuming the average V magnitude for the RR Lyrae stars
 of  $<$V$>$$\sim$16.40, and our reddening A$_{\rm V}$=1.165, 
we obtain exactly the same intrinsic distance modulus values
 presented in Table \ref{distance}, of (m-M)$_0$$\sim$14.585. 
This shows that the distances from both methods, that is,
isochrone fit and RR Lyrae stars, are converging. 
The uncertainty in the M$_{\rm V}$ of $\pm$0.10 leads to
 an uncertainty of $\pm$0.10 in (m-M)$_0$, 
and an uncertainty of $^{+0.39}_{-0.37}$ kpc in the distance. 

\begin{table}
\caption{Literature HB level, reddening,
 and distance for NGC~6558. References: 1 Terndrup (1988),
2 Hazen (1996), 3 Rich et al. (1998), 4 Davidge et al. (2004),
5 Barbuy et al. (2007),
6 Rossi et al. (2015), 7 present work. }             
\label{distance}      
\scalefont{0.9}
\centering                          
\begin{tabular}{c@{}c@{}c@{}c@{}c@{}c@{}c@{}cc}        
\hline\hline                 
\noalign{\smallskip}
\hbox{V$_{HB}$} &\hbox{\phantom{-}E(B-V)} &\hbox{\phantom{-}E(V-I)}
 & \hbox{\phantom{-}A$_{V}$} &
\hbox{\phantom{-}(m-M)} &\hbox{\phantom{-}(m-M)$_{o}$}
 &\hbox{\phantom{-}d$_{\odot}$(kpc)}  &\hbox{\phantom{-}reference} & \\
\noalign{\smallskip}
\hline
\noalign{\smallskip}
---     & 0.38    & 0.50   & 1.33    & ---     & ---     &  ---    & 1 & \\
16.00   & 0.43    & ---    & 1.30    & 15.40   & 14.10   &  6.6    & 2 & \\
16.05   & 0.46    & 0.6    & 1.43    & ---     & ---     &  6.4    & 3 & \\
16.30   & 0.44    & ---    & 1.36    & ---     & ---     &  ---    & 4 & \\
15.50   & 0.38    & ---    & 1.18    & ---     & 14.43   &  7.4    & 5 & \\
16.00   & ---     & ---    & 1.18    & ---     & 14.43   &  7.4    & 6 & \\
16.37   & 0.38    & 0.50   & 1.165   & 15.75   & 14.585  &  8.26   & 7 & \\
\noalign{\smallskip}
\hline                                   
\end{tabular}
\end{table}

\section{Spectroscopic observations}

The spectra of individual stars of NGC 6558
 were obtained at the VLT using the UVES spectrograph
(Dekker et al. 2000), in FLAMES-UVES mode. 
 The red chip (5800-6800 {\rm \AA}) uses
ESO CCD \# 20, an MIT backside illuminated, of 4096x2048 pixels, and pixel
size  15x15$\mu$m.
The blue chip (4800-5800 {\rm \AA}) uses ESO Marlene EEV
CCD\#44, backside illuminated, of 4102x2048 pixels, 
and pixel size  15x15$\mu$m. 
The UVES standard setup 580 yields a resolution R $\sim$ 45,000 for a 1
arcsec slit width. 
 The pixel scale is 0.0147 {\rm \AA}/pix, 
with $\sim$7.5 pixels  per resolution element at 6000 {\rm \AA}.


 The data were reduced using the 
UVES pipeline, within ESO/Reflex software (Ballester et al. 2000;
Modigliani et al. 2004). 
The log of the 2014 observations are given in  Table \ref{logobs}. 
 The   spectra   were  flatfielded,  optimally extracted   and
wavelength calibrated    with the  FLAMES-UVES pipeline.

The present UVES observations centred at 5800 {\rm \AA} yield
a spectral coverage of 4800 $<$ $\lambda$ $<$ 6800 {\rm \AA},
with a gap at 5708-5825 {\rm \AA}.
The final signal-to-noise ratio  (S/N) resulted in values of
 S/N$\sim$100 in the red portion, 
and S/N$\sim$70 in the blue portion.
The selected stars, their OGLE and 2MASS designations,  coordinates,
 and V, I magnitudes 
 from the NTT 2012 data presented here, 
 together with the 2MASS JHK$_{s}$ (Skrutskie et al. 2006), 
and VVV JHK$_{s}$ magnitudes (Saito et al. 2012), 
are listed in Table~\ref{starmag}.

We derived radial velocities using the IRAF/FXCOR task for
all individual observations in each run, adopting as template
the Arcturus spectrum (Hinkle et al. 2000).
The present mean heliocentric radial velocity v$^{\rm hel}_{\rm r}$ =
 $-$194.45$\pm$2.1 km~s$^{\rm {-1}}$ is in very good agreement
with the value of  v$^{\rm hel}_{\rm r}$ =
 $-$197.3$\pm$4.0 km~s$^{\rm {-1}}$ derived from five giants stars
observed with the GIRAFFE spectrograph (Barbuy et al. 2007),
and  $-$198.0$\pm$4.0 km~s$^{\rm {-1}}$ derived by Saviane et al. (2012)
from four stars. 
Spectra  extracted  from different frames were  then  co-added, taking into 
account the radial velocities reported in Table \ref{vr}.

   \begin{table*}
\caption{Log of the spectroscopic observations: date, time,
Julian date, exposure time, 
seeing, and air mass at the beginning and at the end of the observation.}             
\label{logobs}      
\scalefont{1.0}
\centering                          
\begin{tabular}{ccccccccc}        
\hline\hline                 
\noalign{\smallskip}
\hbox{Run} & \hbox{Date} & \hbox{Time} & HJD &\hbox{Exp. (s)} &\hbox{Seeing (``)} &\hbox{Airmass}& \\
\noalign{\smallskip}
\hline
\noalign{\smallskip}
\hbox{1}  & 16.04.14 & 07:33:30 & 2456763.81493  & 2700 & 1.09$-$1.03 & 0.71$-$0.66 &\\
\hbox{2}  & 16.04.14 & 08:21:18 & 2456763.84812  & 2700 & 1.03$-$1.01 & 0.66$-$1.07 &\\
\hbox{3}  & 16.04.14 & 09:10:32 & 2456763.88231 & 2700 & 1.01$-$1.02 & 1.02$-$1.07 &\\
\hbox{4}  & 29.04.14 & 05:30:49 & 2456776.72973  & 2700 & 1.00$-$1.13 & 1.28$-$1.14 &\\
\hbox{5}  & 07.06.14 & 05:59:48 & 2456815.74986  & 2700 & 0.91$-$0.91 & 1.01$-$1.03 &\\
\hbox{6}  & 04.07.14 & 02:50:40 & 2456842.61852 & 2700 & 1.46$-$1.00 & 1.05$-$1.01 &\\
\noalign{\smallskip}
\hline                                   
\end{tabular}
\end{table*}

\begin{table*}
\caption[1]{Identifications, positions, and magnitudes.
$JHK_{s}$ from both 2MASS and VVV surveys are given.
Superscript 1: uncertain values. }
\small
\begin{flushleft}
\tabcolsep 0.15cm
\begin{tabular}{ccccccccccccccccccc}
\noalign{\smallskip}
\hline
\noalign{\smallskip}
\hline
\noalign{\smallskip}
{\rm  OGLE no.}& 2MASS ID & $\alpha_{2000}$ & $\delta_{2000}$ & $V$ & $I$ & $J$ & $H$ & $K_{\rm s}$ &   $J_{\rm VVV}$ 
& {\rm H$_{\rm VVV}$} & {\rm K$_{\rm VVV}$} &  \cr
\noalign{\vskip 0.05cm}
\noalign{\hrule}
\noalign{\vskip 0.05cm}
283 & 18102223-3145435 &  18 10 22.228 &  -31 45 43.340     &  15.883 & 14.378 & 13.248 &12.564 &10.560$^{1}$   &13.235 & 12.625 &12.502 & \\
 364 & 18102150-3145268 & 18 10 21.496 & -31 45 26.820      &  15.738 & 14.203 & 13.128 &12.449 &12.316   &13.047 & 12.462 &12.313 & \\
1072 & 18101520-3146014 & 18 10 15.208 & -31 46  1.370     &  15.960 &  14.345 & 13.183 &12.481 &12.338  &--- & 12.503 &12.377 & \\
 1160 &18101768-3145246 &  18 10 17.682 & -31 45 24.570     &  15.586 & 14.063 &12.892  &12.000 &11.863   &12.902 & 12.320 &12.196 & \\
 \hline
\end{tabular}
\end{flushleft}
\label{starmag}
\end{table*}

\begin{table}
\caption[2]{Radial velocities of the UVES sample stars, in each of the six
exposure runs, corresponding heliocentric radial velocities and mean
heliocentric radial velocity.}
\small
\begin{flushleft}
\begin{tabular}{l@{}c@{}c@{}c@{}c@{}c@{}c@{}c@{}}
\noalign{\smallskip}
\hline
\noalign{\smallskip}
\hline
\noalign{\smallskip}
Target & \phantom{-}${\rm v_r^{obs}}$ & \phantom{-}\phantom{-}${\rm v_r^{hel.}}$ & & \phantom{-}Target & \phantom{-}${\rm v_r^{obs}}$ & 
\phantom{-}\phantom{-}${\rm v_r^{hel.}}$ \\
\noalign{\smallskip}
&${\rm km~s^{-1}}$ &${\rm km~s^{-1}}$ & & &${\rm km~s^{-1}}$ &${\rm km~s^{-1}}$ \\
\noalign{\smallskip}
\noalign{\smallskip}
\hline
\noalign{\vskip 0.2cm}
1160 1 & \phantom{-}-217.9820 & \phantom{-}-190.76 & & \phantom{-}283 1 & \phantom{-}-221.4165 & \phantom{-}-194.20 & \\ 
1160 2 & \phantom{-}-217.8438 & \phantom{-}-190.72 & & \phantom{-}283 2 & \phantom{-}-221.3267 & \phantom{-}-194.20 & \\ 
1160 3 & \phantom{-}-218.2692 & \phantom{-}-191.23 & & \phantom{-}283 3 & \phantom{-}-221.0840 & \phantom{-}-194.05 & \\ 
1160 4 & \phantom{-}-214.5870 & \phantom{-}-190.52 & & \phantom{-}283 4 & \phantom{-}-217.4875 & \phantom{-}-193.42 & \\ 
1160 5 & \phantom{-}-198.9683 & \phantom{-}-190.79 & & \phantom{-}283 5 & \phantom{-}-202.3398 & \phantom{-}-194.16 & \\ 
1160 6 & \phantom{-}-185.4816 & \phantom{-}-190.23 & & \phantom{-}283 6 & \phantom{-}-189.3486 & \phantom{-}-194.10 & \\ 
\hbox{Mean} & --- & \phantom{-}-190.70 & & \hbox{Mean} & --- & \phantom{-}-194.02 & \\
\noalign{\smallskip}
\hline
\noalign{\vskip 0.2cm}
364 1 & \phantom{-}-222.1566 & \phantom{-}-194.94 & & \phantom{-}1072 1 & \phantom{-}-226.2227 & \phantom{-}-199.00 & \\
364 2 & \phantom{-}-222.0449 & \phantom{-}-194.92 & & \phantom{-}1072 2 & \phantom{-}-226.2266 & \phantom{-}-199.10 & \\
364 3 & \phantom{-}-221.5339 & \phantom{-}-194.50 & & \phantom{-}1072 3 & \phantom{-}-226.3010 & \phantom{-}-199.26 & \\
364 4 & \phantom{-}-217.9897 & \phantom{-}-193.93 & & \phantom{-}1072 4 & \phantom{-}-222.1284 & \phantom{-}-198.06 & \\
364 5 & \phantom{-}-202.4231 & \phantom{-}-194.25 & & \phantom{-}1072 5 & \phantom{-}-206.6227 & \phantom{-}-198.45 & \\
364 6 & \phantom{-}-189.6272 & \phantom{-}-194.38 & & \phantom{-}1072 6 & \phantom{-}-193.0282 & \phantom{-}-197.78 & \\

\hbox{Mean} & & \phantom{-}-194.48 & &\hbox{Mean} & --- & \phantom{-}-198.60 & \\
\noalign{\smallskip}
\hline

\noalign{\smallskip}
\hline
\end{tabular}
\end{flushleft}
\label{vr}
\end{table}


\section{Photometric stellar parameters}

\subsection{Temperatures}

$VIJHK_{\rm s}$ magnitudes are given in Table \ref{starmag}.
$V$ and $I$ magnitudes from the NTT 2012 observations,
 2MASS   $J$, $H$, and $K_{\rm s}$   from Skrutskie et al. (2006)\footnote{
$\mathtt{http://ipac.caltech.edu/2mass/releases/allsky/}$} and  VVV $J$, $H$, and $K_{\rm s}$ from the VVV survey
(Saito et al. 2012), were adopted. Reddening laws from 
Dean et al.   (1978) and  Rieke  \& Lebofsky (1985), namely,  
R$_{\rm V}$ = A$_{\rm V}$/E($B-V$) = 3.1,
 E($V-I$)/E($B-V$)=1.33,
 E($V-K$)/E($B-V$)=2.744,
E($J-K$)/E($B-V$)=0.527 were used.

Effective temperatures were derived from  $V-I$ using
the colour-temperature calibrations of Alonso et al.  (1999).
 For the transformation of $V-I$ from Cousins to Johnson system
 we used $V-I$$_{C}$=0.778($V-I$$_{J}$) as given in  Bessell (1979).
The VVV $JHK_{s}$ colours were transformed to the 2MASS $JHK_{s}$ system, using
relations by Soto et al. (2013). For 
the effective temperatures derived from
the  $V-K$, and $J-K$ 2MASS and VVV colours we used the relations by
Gonzalez Hernandez \& Bonifacio (2009). It seems that some magnitudes are not
adequate, in particular the 2MASS $K_{s}$ for star 283.
 The photometric effective
temperatures derived are  listed in  Table~\ref{tabteff}. 

\subsection{Gravities}

 The following  classical  relation was used to derive gravities

\begin{equation}
\label{eq1}
\log g_*=4.44+4\log \frac{T_*}{T_{\odot}}+0.4(M_{\rm bol*}-M_{\rm bol{\circ}})+\log \frac{M_*}{M_{\odot}} 
\end{equation}

For NGC 6558 the NTT 2012 photometry gives A$_V$ = 1.17, and
(m-M)$_{\circ}$ = 14.58.
We adopted T$_{\odot}$=5770 K, M$_{\rm bol \odot}$=4.75, and M$_*$=0.85 M$_{\odot}$.
 The   bolometric  corrections computed with relations from Alonso
 et al. (1999), and
corresponding gravities are given in Table~\ref{tabteff}.
The photometric parameters were adopted as an initial guess in
order to proceed with the analysis of the spectroscopic data for the derivation
of the final stellar parameters, as described in the next section. 





\section{Spectroscopic stellar parameters}

The  equivalent  widths (EW) were initially
 measured  using  the  automatic  code
DAOSPEC, developed by Stetson \& Pancino (2008).
We also measured all lines manually using IRAF.
 The \ion{Fe}{II} lines adopted were all measured by hand
using IRAF, given that DAOSPEC is unable to take subtle blends into account.
To rely on the \ion{Fe}{II} lines to derive the gravities,
we compared the observed and synthetic lines, and eliminated the
not clearly present lines in the observed spectra. 
For the \ion{Fe}{I} and \ion{Fe}{II}  lines, 
the oscillator strengths and EWs are reported in Table to be
published electronically only.
The wavelengths, excitation potentials and oscillator strengths
were gathered from the line lists by Kurucz (1993) 
websites\footnote{http://www.cfa.harvard.edu/amp/amp\-data/ku\-rucz23/\-se\-kur.\-html}\footnote{http://kurucz.harvard.edu/atoms.html}, the
National Institute of Standards \& Technology (NIST, Martin et al. 2002)
\footnote{http://physics.nist.gov/PhysRefData/ASD/lines$_-$form.html}, 
and VALD (Piskunov et al. 1995).
  Oscillator strengths for the \ion{Fe}{II} lines are
from Mel\'endez \& Barbuy (2009).

 Atomic constants (log~gf and van der Waals damping constants,
and hyperfine structure) were adopted
after checking the fits to lines in the solar spectrum
with the same UVES@VLT instrumentation as the present data
\footnote{www.eso.org/observing/dfo/quality/UVES/pipeline/\-so\-lar\-$_{--}$spec\-trum.\-html}, the
Arcturus spectrum (Hinkle et al. 2000), and the metal-rich star
$\mu$ Leo (Lecureur et al. 2007), as explained in previous papers
(Barbuy et al. 2006; Barbuy et al. 2014; Trevisan \& Barbuy 2014, 
among others).

Photospheric 1-D models for the sample giants  were extracted from the
MARCS model atmospheres grid (Gustafsson   et al. 2008).
The LTE  abundance  analysis and  the spectrum  synthesis calculations
were performed using the code described in Barbuy et al. (2003), 
Coelho et al. (2005),
and Barbuy et al. (2018b). A solar iron abundance
of $\epsilon$(Fe)=7.50 (Grevesse \& Sauval 1998) was adopted.
Molecular lines
of CN  (A$^2$$\Pi$-X$^2$$\Sigma$), C$_2$ Swan (A$^3$$\Pi$-X$^3$$\Pi$), TiO
(A$^3$$\Phi$-X$^3$$\Delta$) $\gamma$   and  TiO (B$^3$$\Pi$-X$^3$$\Delta$)
$\gamma$' systems  are taken   into  account in the studied wavelength region.

The stellar   parameters  were  derived    by initially  adopting  the
photometric effective  temperature  and  gravity,  and   then  further
constraining  the  temperature by imposing  excitation equilibrium for
\ion{Fe}{I} lines.
The selected \ion{Fe}{II} lines measurable,
   allowed to derive gravities imposing ionization equilibrium, noting
that their EWs are uncertain due to continuum placement inaccuracy.
For this reason, we have over-plotted 
the computed and observed \ion{Fe}{II} lines, 
and selected only those identified to be not blended.
Microturbulence velocities v$_t$  
were  determined by canceling the trend of \ion{Fe}{I} abundance vs.  
equivalent width.

The final spectroscopic parameters T$_{\rm eff}$, log g, [\ion{Fe}{I}/H],
 [\ion{Fe}{II}/H],  [Fe/H] and
 v$_t$ values  are reported in  the  last columns
  of Table~\ref{tabteff}. 
We have adopted the spectroscopic parameters for the derivation of
elemental abundances.

\begin{figure}
\centering
\includegraphics[scale=0.37]{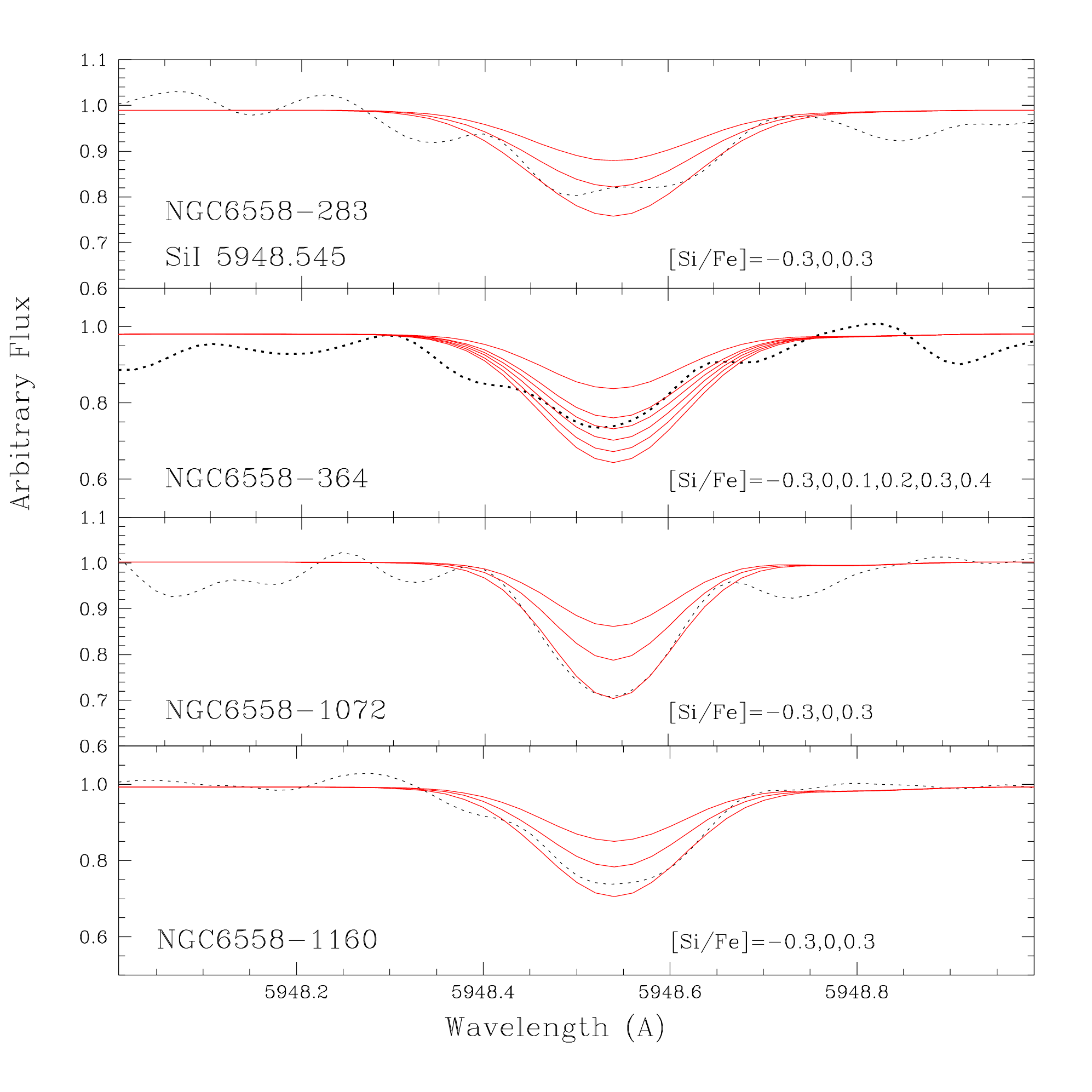}
\includegraphics[scale=0.37]{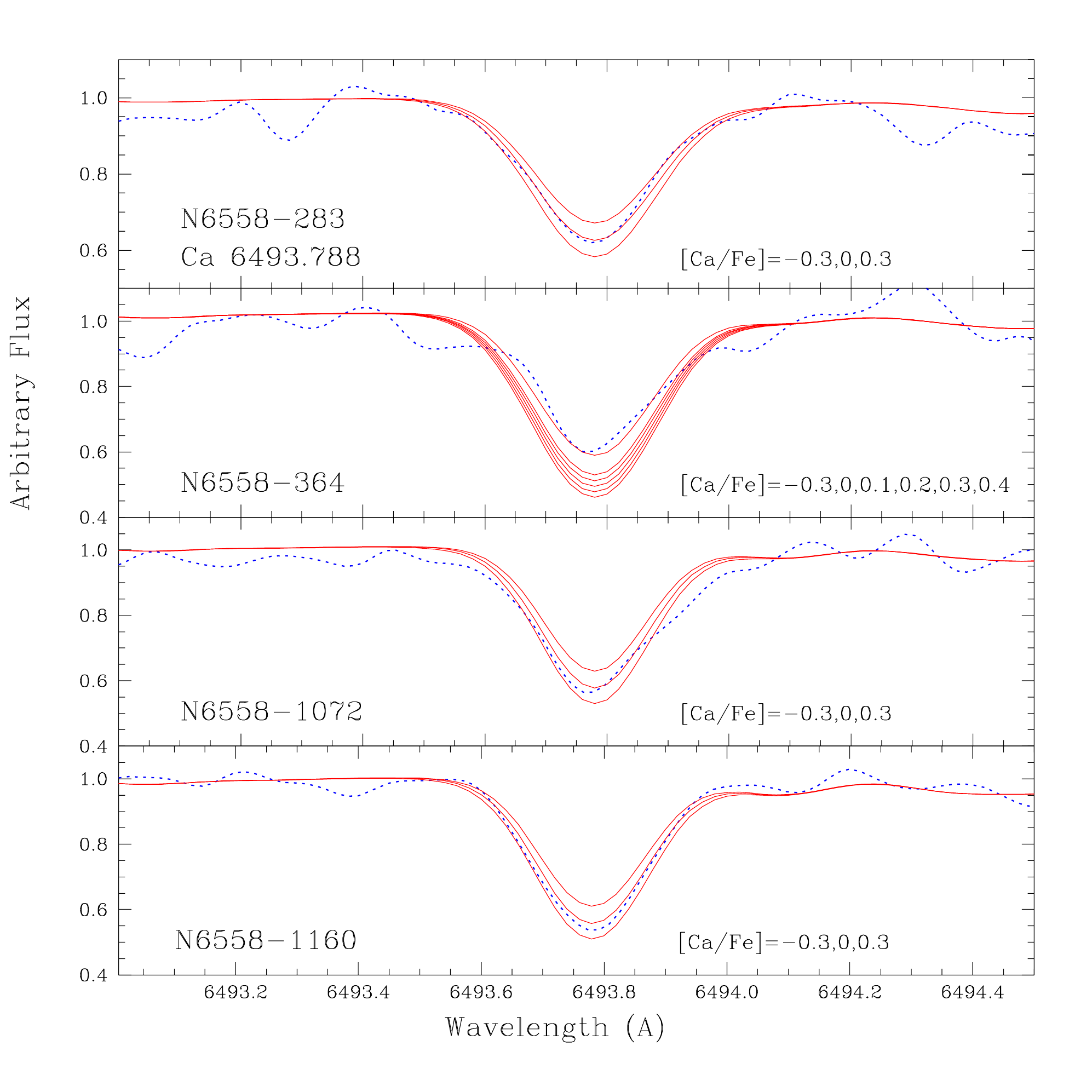}
\includegraphics[scale=0.37]{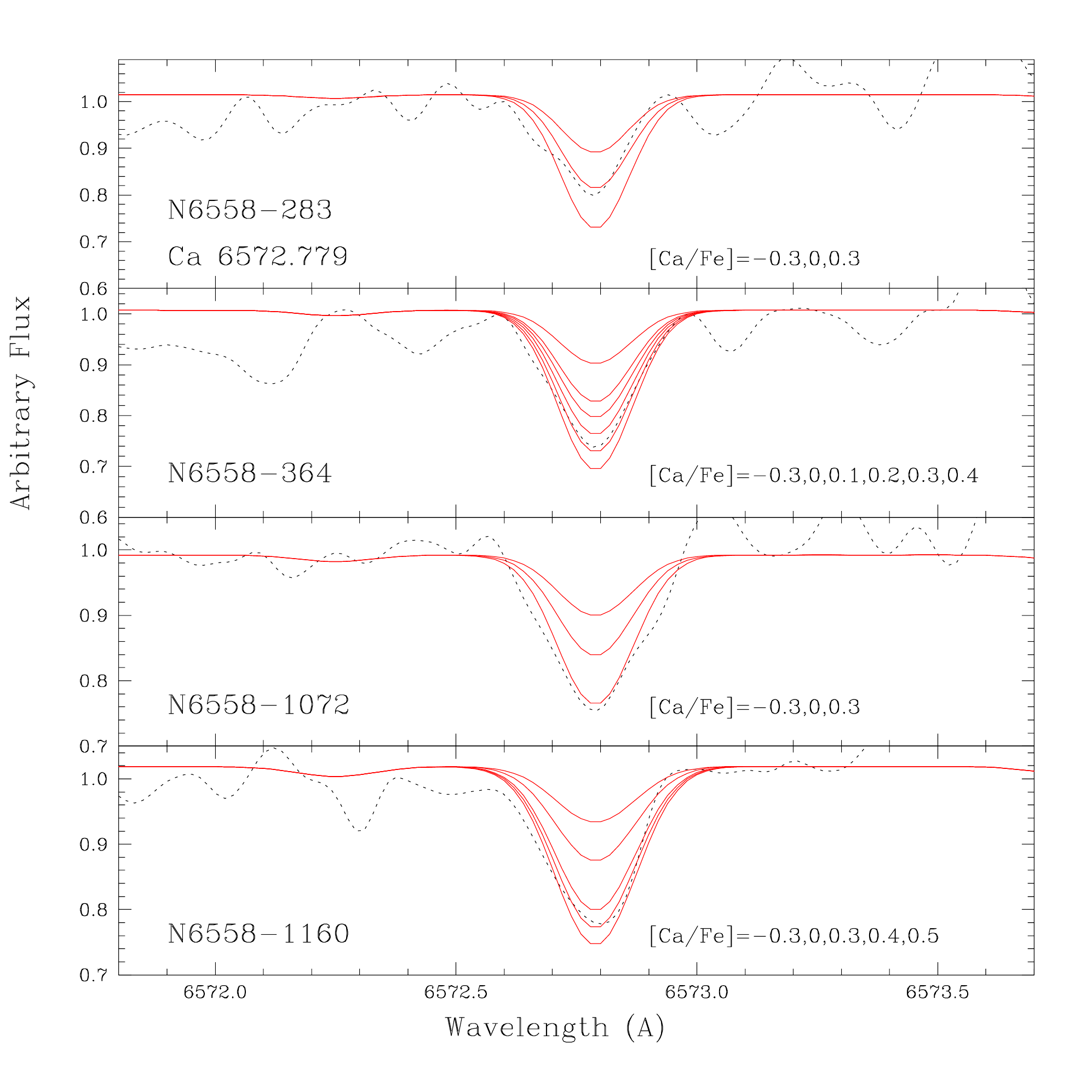}
\caption{SiI 5948.545, CaI 6493.788 and CaI 6572.779 {\rm \AA} lines
in the four sample stars. Calculations for different [Si/Fe] and [Ca/Fe]
are shown, with values as indicated in each panel.
}
\label{sica} 
\end{figure}

\begin{figure}
\centering
\includegraphics[scale=0.37]{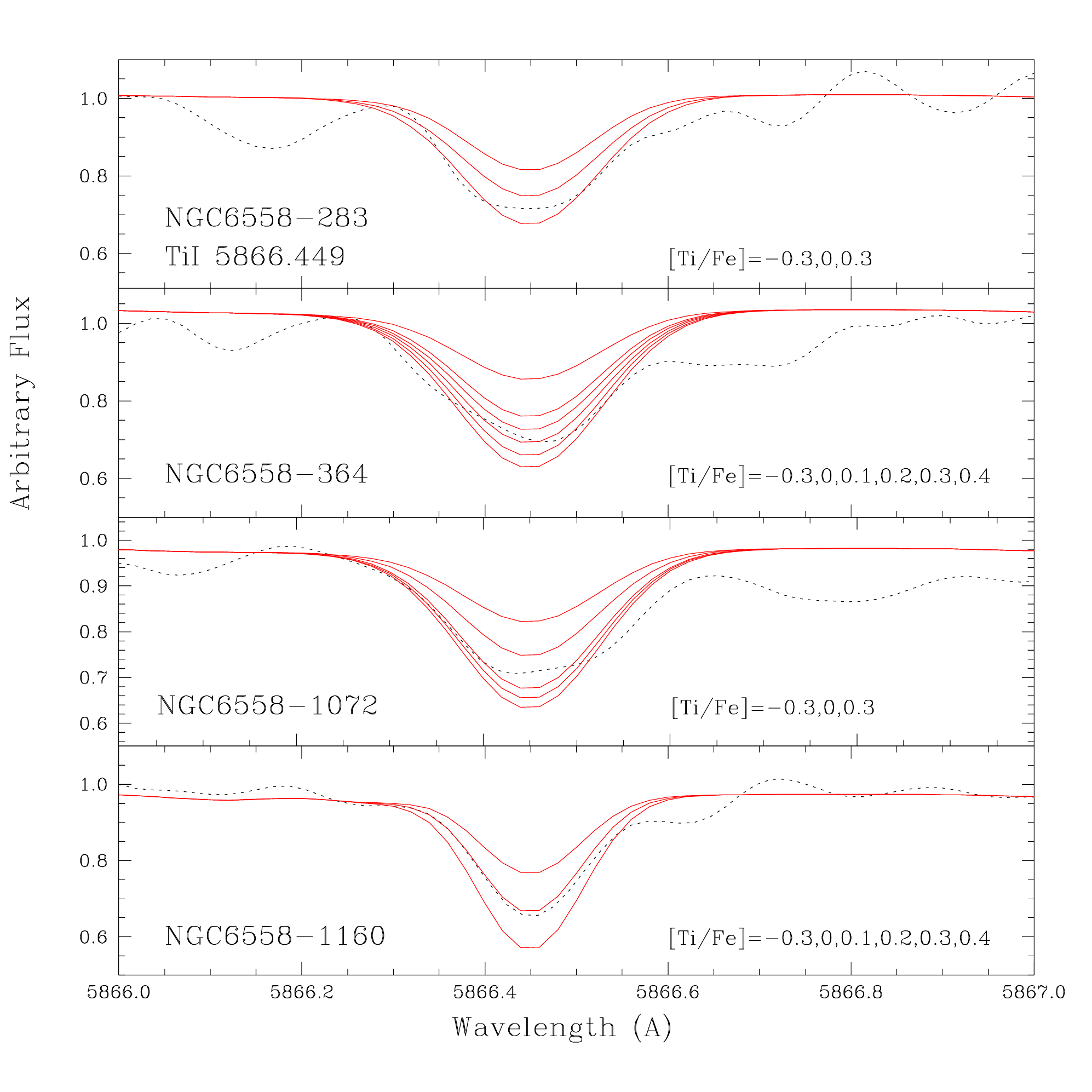}
\includegraphics[scale=0.37]{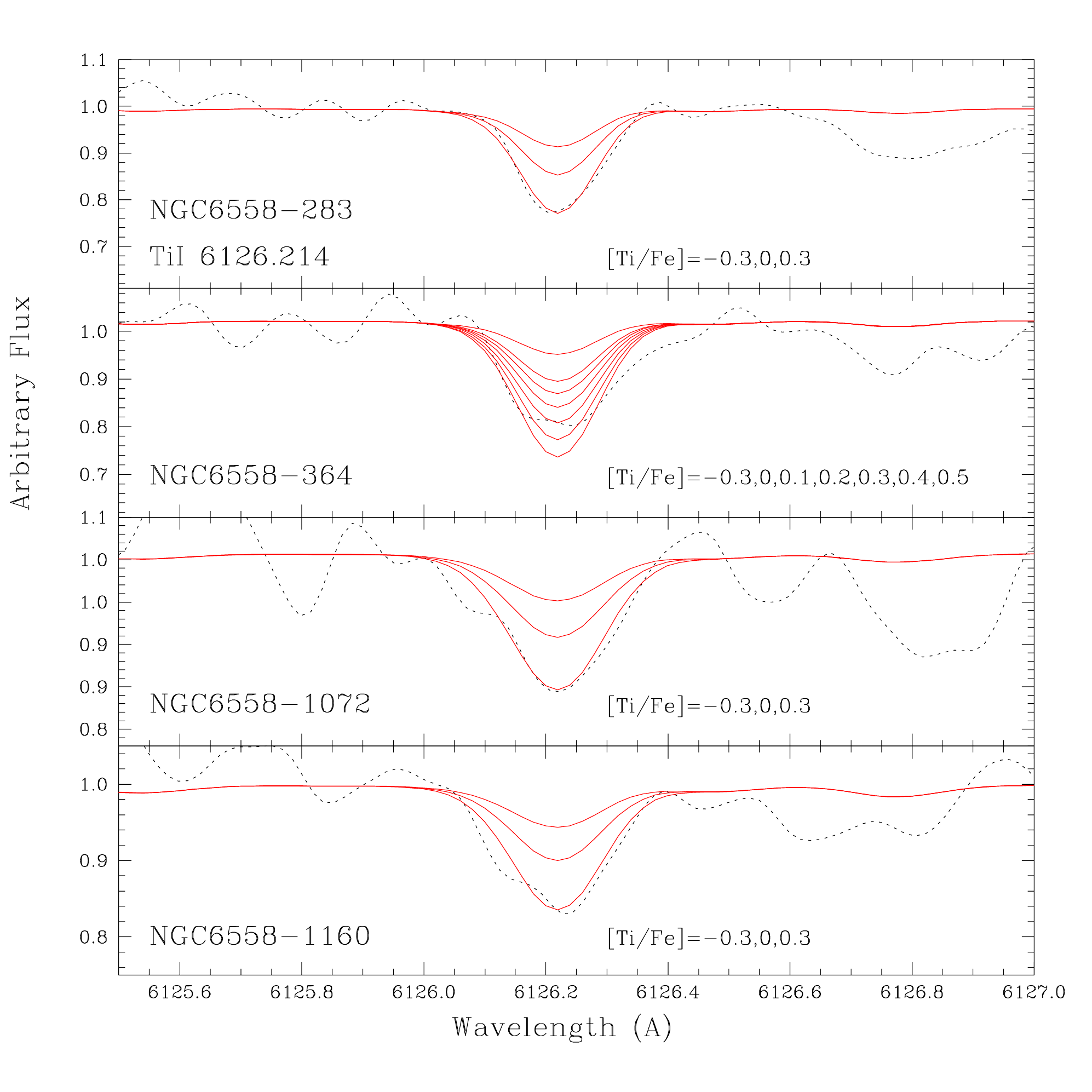}
\caption{TiI 5866.449 and 6126.214  {\rm \AA} lines
in the four sample stars. Calculations for different [Ti/Fe]
are shown, with values as indicated in each panel.
}
\label{ti} 
\end{figure}

\begin{figure}
\centering
\includegraphics[scale=0.37]{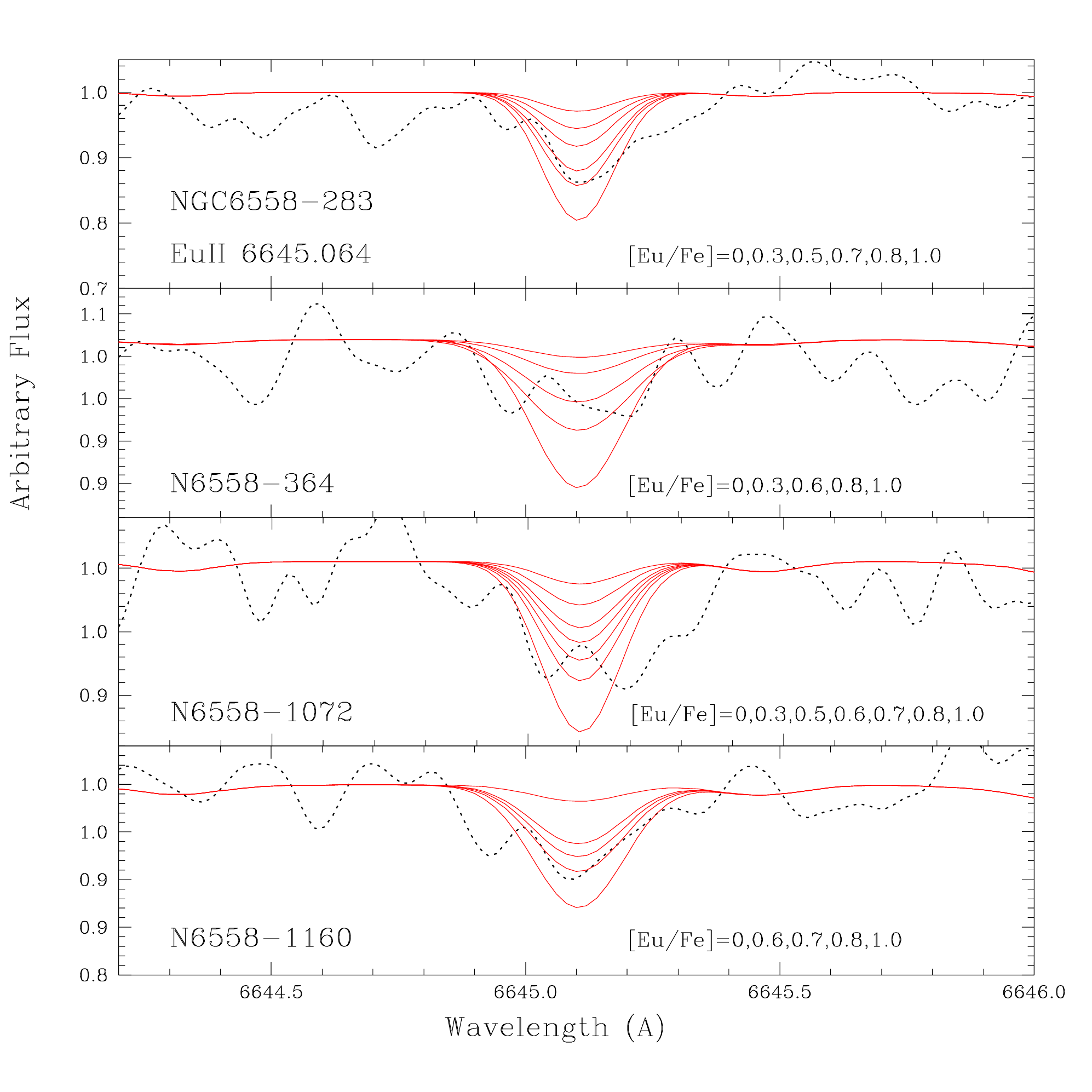}
\includegraphics[scale=0.37]{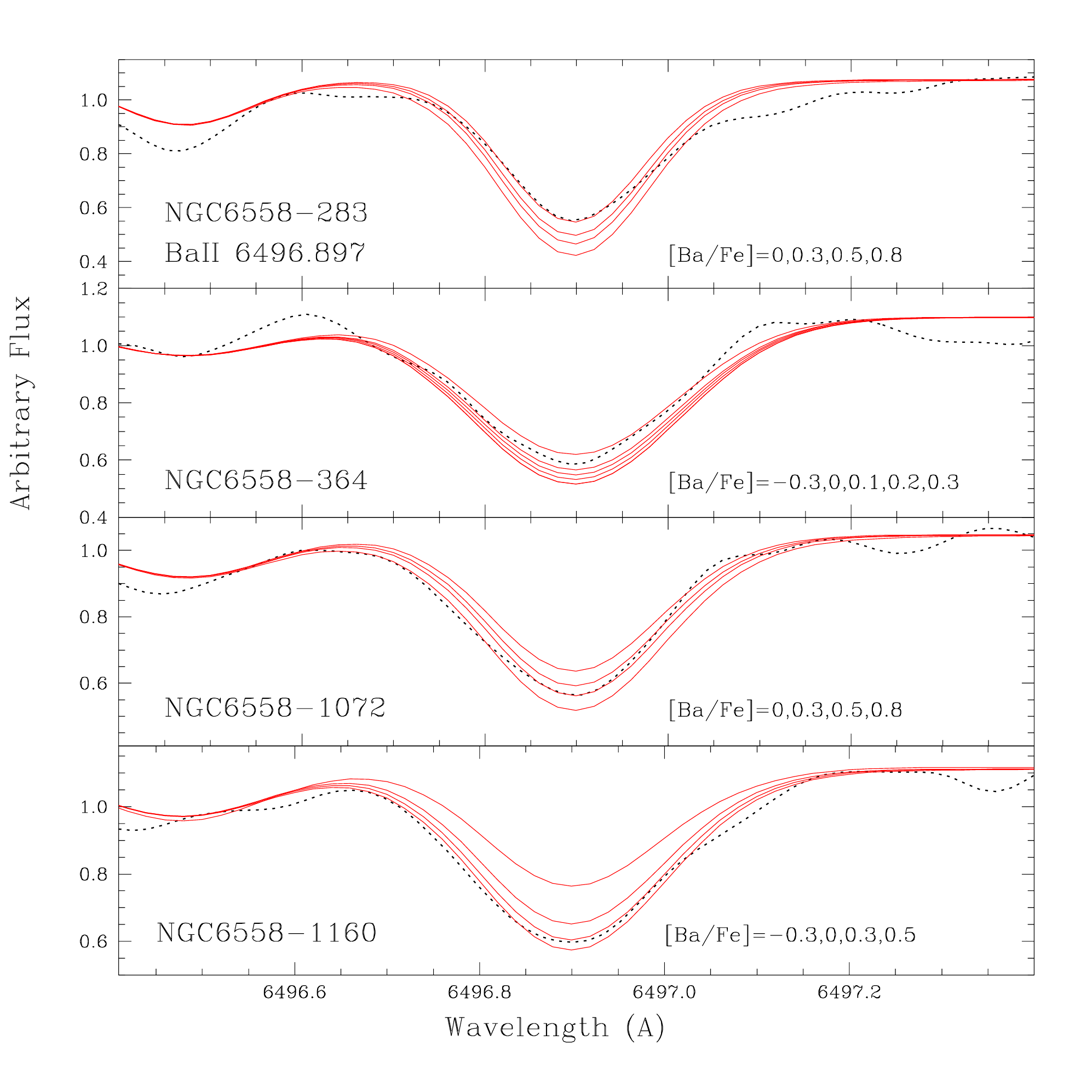}
\caption{EuII 6645.064 and BaII 6496.897  {\rm \AA} lines
in the four sample stars. Calculations for different [Eu/Fe] and [Ba/Fe]
are shown, with values as indicated in each panel.
}
\label{euba} 
\end{figure}

\begin{figure}
\centering
\includegraphics[scale=0.45]{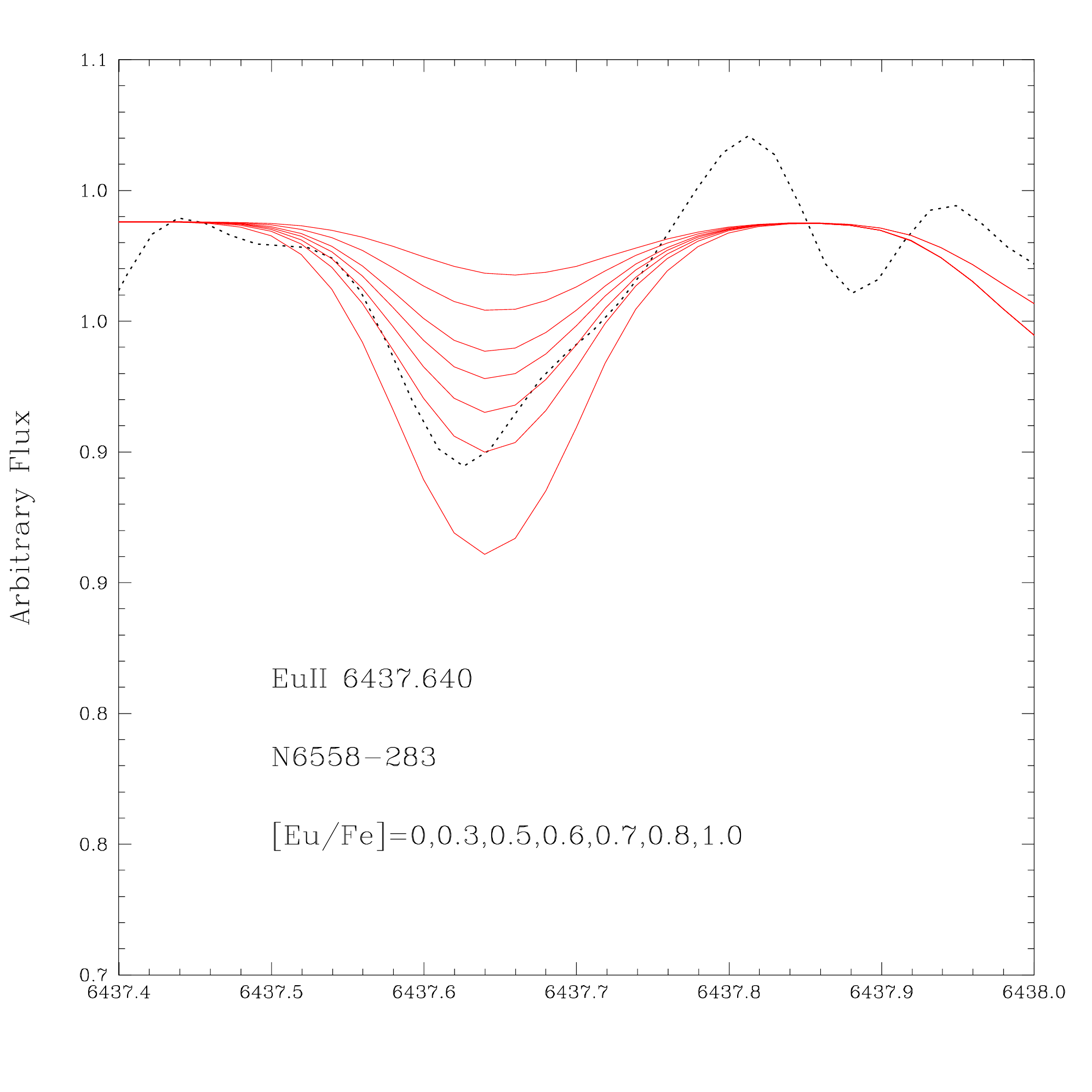}
\caption{EuII 6347.640  {\rm \AA} in star 283.
}
\label{eu64} 
\end{figure}

\begin{figure}
\centering
\includegraphics[scale=0.45]{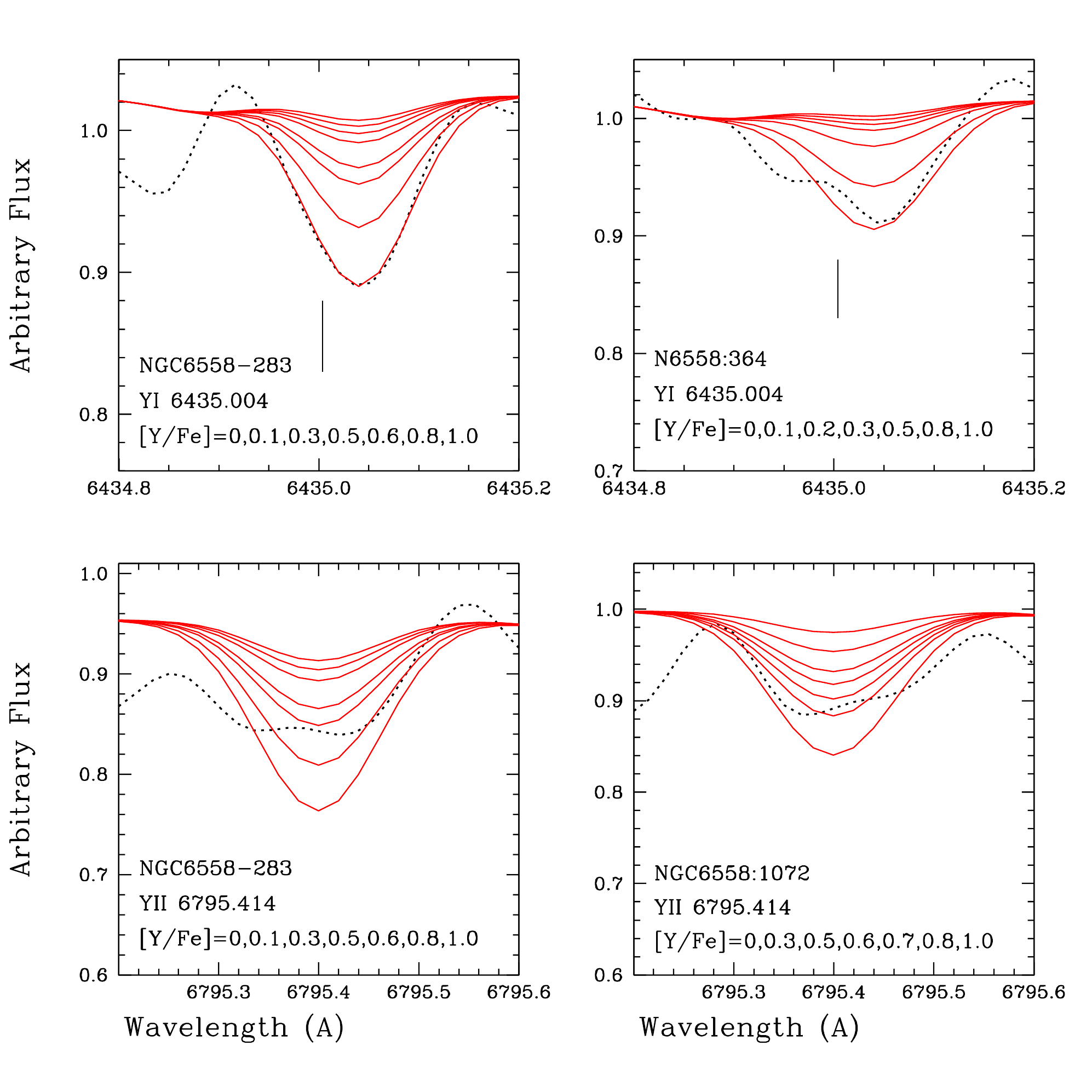}
\caption{YI 6435.004 and YII 6795.414  {\rm \AA} lines
in stars  283, 364 and 1072. Calculations for different [Y/Fe]
are shown, with values as indicated in each panel.
}
\label{y} 
\end{figure}

\begin{figure}
\centering
\includegraphics[scale=0.45]{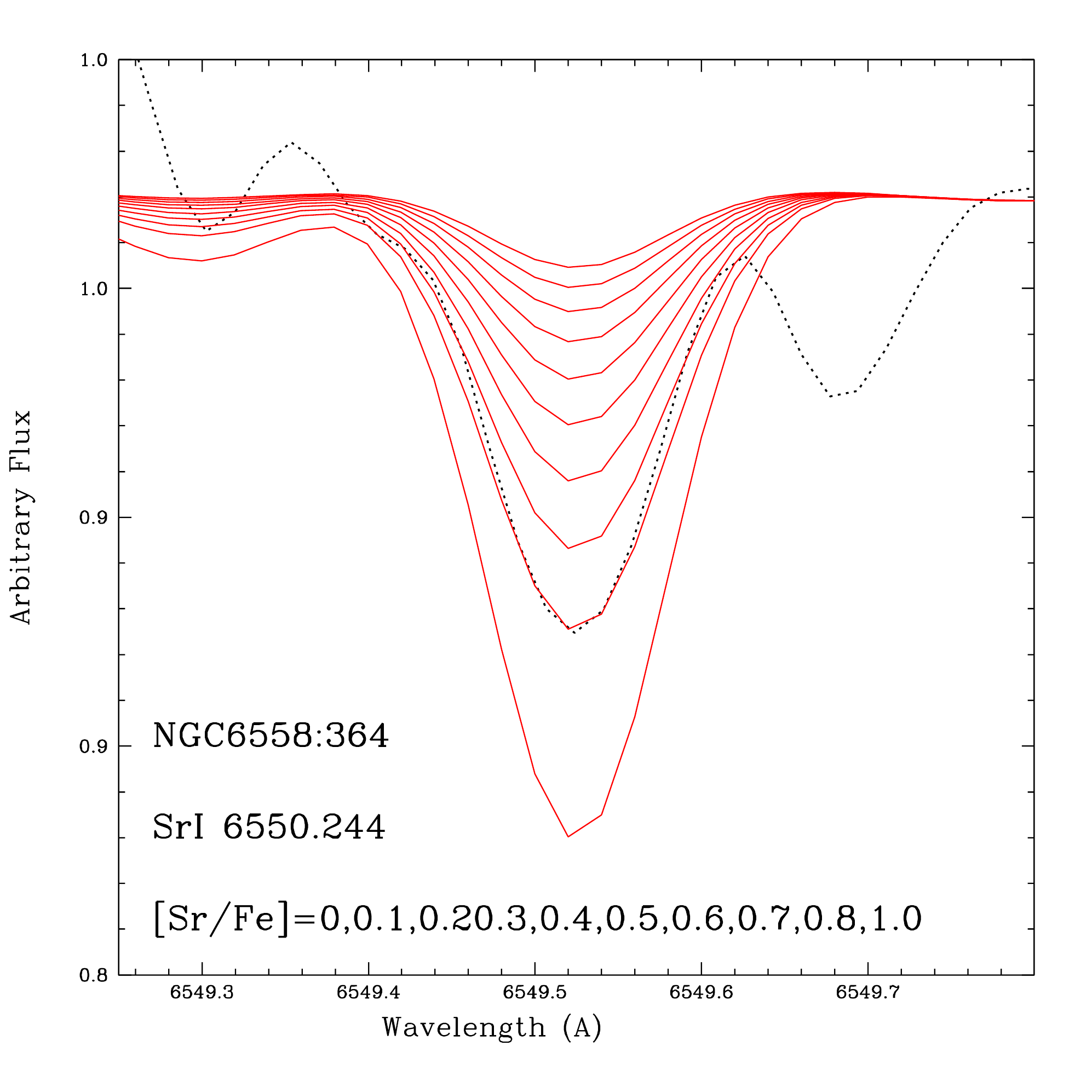}
\caption{SrI 6550.244 {\rm \AA} line in star 364.
Calculations for different values of [Sr/Fe]
as indicated in the panel are shown.
}
\label{sr} 
\end{figure}

\begin{table*}
\caption[1]{
Photometric stellar parameters derived using the calibrations by Alonso et al. (1999)
 for $V-I$, $V-K$, $J-K$, bolometric corrections, bolometric magnitudes
and corresponding gravity log $g$,
and final spectroscopic parameters.} 
\small
\begin{flushleft}
\begin{tabular}{c@{}c@{}c@{}c@{}c@{}c@{}c@{}c@{}cc@{}c@{}c@{}c@{}c@{}c@{}c}
\noalign{\smallskip}
\hline
\noalign{\smallskip}
\hline
\noalign{\smallskip}
& \multicolumn{4}{c}{\hbox{}} Photometric\phantom{-} parameters & & & \multicolumn{7}{c}{\hbox{}} Spectroscopic\phantom{-} parameters\\
\cline{2-8}  \cline{10-15}    \\
{\rm star} & T($V-I$) & \phantom{-}\phantom{-}T($V-K$) & \phantom{-}\phantom{-}T($J-K$)
 &\phantom{-}\phantom{-}T($V-K$)  &  \phantom{-}\phantom{-}T($J-K$) &  
\phantom{-}\phantom{-}${\rm BC_{V}}$ & 
\phantom{-}\phantom{-}log g & & \phantom{-}T$_{\rm eff}$ &
\phantom{-}\phantom{-}log g &\phantom{-}\phantom{-}[FeI/H] & 
\phantom{-}\phantom{-}[FeII/H] &
 \phantom{-}\phantom{-}[Fe/H] 
& ${\rm v_t }$ \\
 &  & 2MASS & 2MASS  & VVV  &VVV  &   & & &  &  &    \\
 & (K) & (K) &  (K) &  (K) & (K) & & &  & (K) & & & & & (km~s$^{-1}$)  \\
\noalign{\smallskip}
\noalign{\hrule}
\noalign{\smallskip}
    283  &  4730 &  3647 &  2103 &  4823 & 4895 & $-0.36$  & \phantom{-}2.1 & & 4800 & \phantom{-}2.10 & $-$1.20  & $-$1.20 & $-$1.20 & 1.00 \\
    364  &  4666 &  4792 &  4831 &  4784 & 4899 & $-0.39$  & \phantom{-}2.0 & & 4880 & \phantom{-}2.20 & $-$1.18 & $-$1.13 & $-$1.15 & 1.90 \\
    1072 &  4509 &  4621 &  4728 &  4020 & ---   & $-0.48$  & \phantom{-}2.0 & & 4850 & \phantom{-}2.60 & $-$1.17  & $-$1.17 & $-$1.17 & 1.01 \\
    1160 &  4692 &  4541 &  4222 &  4815 & 4994    & $-0.38$  & \phantom{-}2.0 & & 4900 & \phantom{-}2.60 & $-$1.11  & $-$1.18 & $-$1.15 & 1.30 \\
\noalign{\smallskip} \hline \end{tabular}
\end{flushleft}
\label{tabteff}
\end{table*}




\section{Element abundances}

Abundances  were   obtained  by means of line-by-line  spectrum
synthesis calculations compared with the observed spectra.
The hyperfine structure (HFS) for the studied lines of 
 \ion{Ba}{II} and \ion{Eu}{II} were taken into account.

We report the line-by-line abundances for the light elements C, N, O, the
 odd-Z elements Na, Al, the  $\alpha$-elements Mg, Si, Ca, Ti, and the heavy elements
Ba, Y, Sr, and Eu abundances, given electronically only.
In Table \ref{final} we give the final abundances for each star,
and the mean abundances for the cluster.

The carbon abundances were estimated from the C$_{2}$(0,1)  bandhead
at 5635.3 {\rm \AA} of the Swan d$^{3}\Pi_g$-a$^{3}\Pi_u$ system,
using the laboratory line list from Phillips \&
Davis (1968). The N abundances were derived from the
CN (5,1) bandhead at 6332.2 {\rm \AA} of the CN 
A$^{2}\Pi$-X$^{2}\Sigma$ red system, using laboratory line lists
by Davis \& Phillips (1963). Both these bandheads
are weak and indicate only an upper limit for the S/N of these
spectra.
  The forbidden [OI] 6300.311 {\rm \AA} line 
was used to derive the oxygen abundances, taking into account
the estimated C and N abundances iteratively.
The abundance derivation of the elements Na, Al, Mg, Si, Ca, and Ti
followed the usual procedures and line lists (e.g. Barbuy et al. 2016).
Examples of fits of synthetic spectra to observed lines are given
in Figures \ref{sica} and \ref{ti}.

Fits to the \ion{Eu}{II} 6645.064 and \ion{Ba}{II} 6496.897 {\rm \AA}
lines are shown in Fig. \ref{euba}. Eu was derived from
the \ion{Eu}{II} 6645.064 {\rm \AA}, showing all high values.
The lines \ion{Eu}{II} 6173.029 {\rm \AA}, \ion{Eu}{II} 6437.640 {\rm \AA}
were also inspected. The former is blended and would in any case
indicate an even higher Eu abundance. The latter is compatible with
the abundances obtained from the \ion{Eu}{II} 6645.064 {\rm \AA} line,
as can be seen in Fig. \ref{eu64} in star 283 (for the other three stars
there are defects in the spectra).

 The heavy element Y was derived only for three stars from the
\ion{Y}{II} 6795.414 {\rm \AA} line, and also from \ion{Y}{II} 6435.004
{\rm \AA} for one star; Sr was derived for one star from one line.
As can be seen from Figs. \ref{y} and \ref{sr} these results must be taken with caution. 
 Finally, La remains undefined given that although the \ion{La}{II} 6172.721 {\rm \AA}
line indicates [La/Fe]$\sim$+1.0 for two sample stars, the region shows
 unidentified blends, whereas \ion{La}{II} 6320.276 {\rm \AA} and 
\ion{La}{II} 6390.477 {\rm \AA}  give considerably lower values; 
\ion{La}{II} 6262.287 {\rm \AA} has defects in all four stars;
besides, if La is high this would be in contrast with a low Ba,
and not compatible with the results from Barbuy et al. (2007),
 so that we prefer not to report final values for La.

\begin{table*}
\caption{Mean abundances of odd-Z elements Na, Al,
 $\alpha$-elements O, Mg, Si, Ca, Ti, and the heavy elements Eu, Ba, Y, Sr.  }
\scalefont{1.0}
\begin{flushleft}
\tabcolsep 0.15cm
\begin{tabular}{ccccccccccccccccc}
\noalign{\smallskip}
\hline
\noalign{\smallskip}
\hline
\noalign{\smallskip}
{\rm Star} & [Na/Fe] & [Al/Fe] & [O/Fe] & [Mg/Fe] & [Si/Fe] & [Ca/Fe] & [TiI/Fe] &[TiII/Fe] & [Eu/Fe] &
 [Ba/Fe] & [Y/Fe] &  [Sr/Fe] & \cr
\noalign{\vskip 0.2cm}
\noalign{\hrule\vskip 0.2cm}
\noalign{\vskip 0.2cm}
283 &  +0.15 & +0.30 & +0.40 & --- & --- & +0.00 &  +0.15 & +0.20 & +0.80 & +0.00 &  +0.75 &  --- & \cr
364 & $-$0.33 & +0.23 & +0.20 & +0.30  & $-$0.17 & +0.03 & +0.25  & +0.20 & +0.60 & -0.20  & +0.80 &  +0.80 & \cr
1072 & +0.08 & +0.07 & +0.50 & +0.30  &   +0.31 & +0.10 & +0.27 &  +0.30 & +0.80 & +0.10  & +0.25 &  --- & \cr
1160 & +0.00  & +0.00 & +0.50 & +0.40 & +0.20 & +0.16 & +0.15 & +0.27 & +0.30 & +0.15 & +0.80 & --- & \cr
\noalign{\vskip 0.05cm}
\noalign{\hrule}
\noalign{\vskip 0.05cm}
Mean & $-$0.03   & +0.15 & +0.40 &  +0.33 & +0.11 & +0.07 & +0.21 & +0.24 & +0.63  & +0.01 & +0.56 & +0.80 & \cr
\noalign{\smallskip} \hline \end{tabular}
\label{final}
\end{flushleft}
\end{table*}

{\it Errors:} The errors due to uncertainties in spectroscopic parameters are given 
in Table \ref{errors}, applied to the sample star NGC 6558:283.
We have adopted the usual errors as for similar samples (Barbuy et al. 2014, 2016):
$\pm$100 K in effective temperature, $\pm$0.2 on gravity,
and $\pm$0.2 km~s$^{-1}$ on the microturbulence velocity.
 Errors are computed by computing models with these modified parameters,
 with changes of  $\Delta$T${\rm
eff}$=+100 K, $\Delta$log  g  =+0.2,  $\Delta$v$_{\rm  t}$ =  0.2  km~s$^{-1}$,
and recomputing lines of different elements.
The error given is the abundance difference needed
to reach the adopted abundances.
A total error estimate is given in Column (5) of Table \ref{errors}.
Additionally there is uncertainty in the continuum placement that we estimate
to be of 0.1 dex, and the total final error is
reported in column (6). 
Other uncertainties for similar stars, such as non-LTE effects
 are discussed more 
extensively in Ernandes et al. (2018).


\begin{table}
\caption{Abundance uncertainties for star N6558:283,
 for uncertainties of $\Delta$T$_{\rm eff}$ = 100 K,
$\Delta$log g = 0.2, $\Delta$v$_{\rm t}$ = 0.2 km s$^{-1}$ and
corresponding total error. The errors are to be
added to reach the reported abundances. A continuum uncertainty
leading to an error of 0.1dex is also added in Column (6).
The error in [Y/Fe] comes from +0.15 for line YI 6435
and -0.05 for line YI 6795 {\rm \AA}.} 
\label{errors}
\begin{flushleft}
\small
\tabcolsep 0.15cm
\begin{tabular}{lcccc@{}c@{}}
\noalign{\smallskip}
\hline
\noalign{\smallskip}
\hline
\noalign{\smallskip}
\hbox{Element} & \hbox{$\Delta$T} & \hbox{$\Delta$log $g$} & 
\phantom{-}\hbox{$\Delta$v$_{t}$} & \phantom{-}\hbox{($\sum$x$^{2}$)$^{1/2}$}
& \phantom{-}\hbox{(+continuum)} \\
\hbox{} & \hbox{100 K} & \hbox{0.2 dex} & \hbox{0.2 kms$^{-1}$} & & \\
\hbox{(1)} & \hbox{(2)} & \hbox{(3)} & \hbox{(4)} & \hbox{(5)}  & \hbox{(6)} \\
\noalign{\smallskip}
\hline
\noalign{\smallskip}
\noalign{\hrule\vskip 0.1cm}
\hbox{[FeI/H]}       &  $-$0.10        &+0.01       & +0.05 &\phantom{+}0.11 &\phantom{+}0.15  \\
\hbox{[FeII/H]}      &  +0.10          &  $-$0.07   & +0.04 &\phantom{+}0.13 &\phantom{+}0.16  \\
\hbox{[O/Fe]}        &  +0.00          & +0.05      &+0.00  &\phantom{+}0.05 &\phantom{+}0.11  \\
\hbox{[NaI/Fe]}      &  +0.05          & +0.00      &+0.00  &\phantom{+}0.05 &\phantom{+}0.11  \\
\hbox{[AlI/Fe]}       & +0.06          & +0.00      &+0.00  &\phantom{+}0.06 &\phantom{+}0.12  \\
\hbox{[MgI/Fe]}      &  +0.00          & +0.01      &+0.00  &\phantom{+}0.01 &\phantom{+}0.11  \\
\hbox{[SiI/Fe] }     & +0.03           & +0.00      &+0.00  &\phantom{+}0.03 &\phantom{+}0.10  \\
\hbox{[CaI/Fe]}      &  +0.08          & +0.00      &+0.01  &\phantom{+}0.08 &\phantom{+}0.13  \\
\hbox{[TiI/Fe]}      &  +0.12          & +0.01      &+0.00  &\phantom{+}0.12 &\phantom{+}0.15  \\
\hbox{[TiII/Fe]}     & $-$0.05         & +0.07      &+0.00  &\phantom{+}0.09 &\phantom{+}0.14  \\
\hbox{[YI/Fe]}       & $^{+0.15}_{-0.05}$ &+0.04      &+0.00  &\phantom{+}0.10  &\phantom{+}0.19  \\
\hbox{[BaII/Fe]}     &  +0.00          & +0.02      &+0.00  &\phantom{+}0.02 &\phantom{+}0.10  \\
\hbox{[EuII/Fe]}     &  $-$0.05        & +0.05      &+0.00  &\phantom{+}0.07 &\phantom{+}0.12  \\
\noalign{\smallskip} 
\hline 
\end{tabular}
\end{flushleft}
 \end{table}


\section{Discussion}

We  derived  a mean metallicity of [Fe/H]=$-$1.17$\pm$0.10 for NGC~6558, 
about 0.2 dex below the results from Barbuy et al. (2007) 
from lower resolution GIRAFFE spectra.
Uncertainties in the continuum placement in the present spectra,
in particular in the measuremente of \ion{Fe}{II} lines, mean that
this metallicity could be slightly underestimated. 
 The UVES spectra, having higher resolutions, should be more accurate than
the GIRAFFE ones, given that resolution is more important than S/N
(Cayrel 1988). Even so there are difficulties in
reliably measuring the \ion{Fe}{II} lines.

\begin{figure*}
\centering
\includegraphics[scale=0.8]{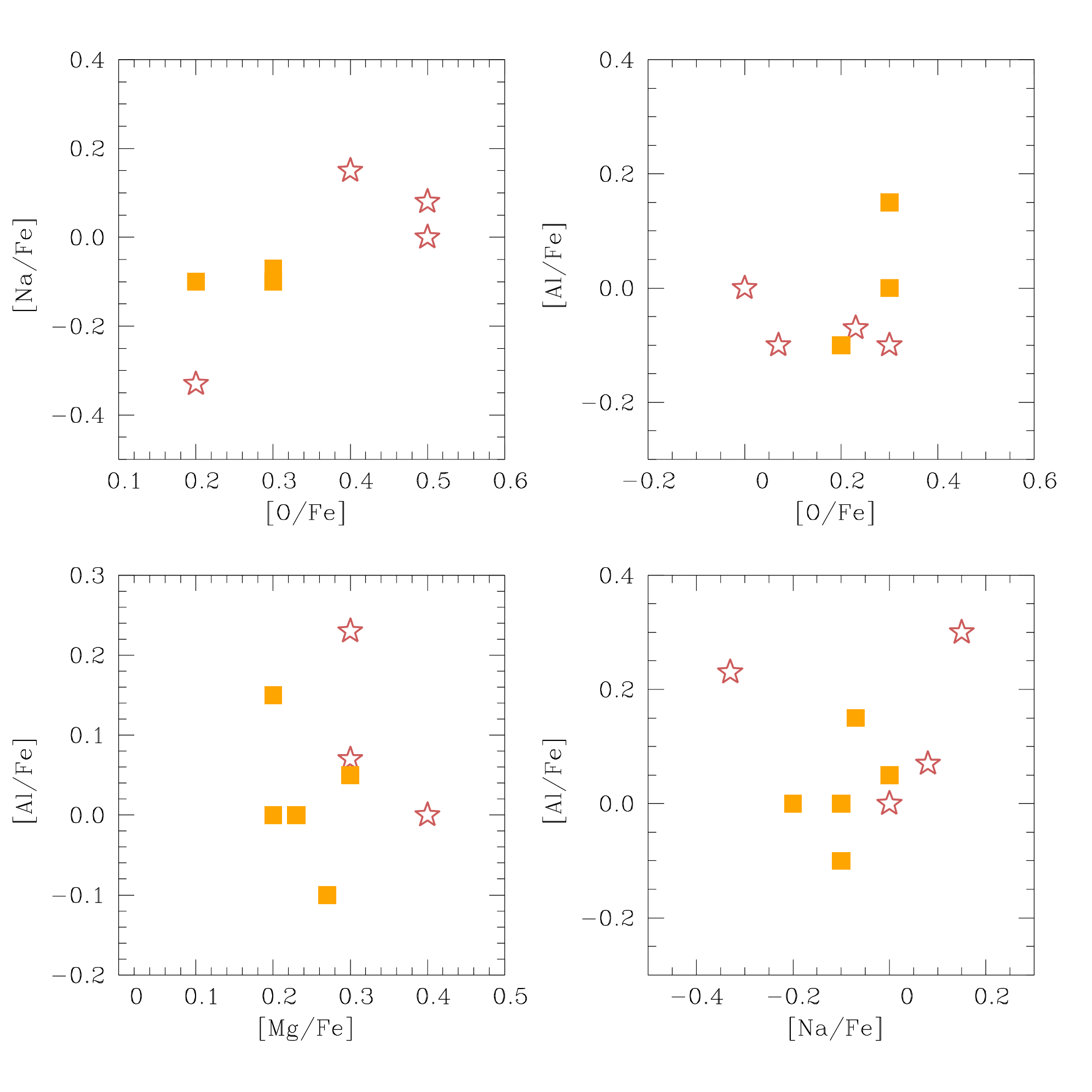}
\caption{[Na/Fe] vs. [O/Fe],  [Al/Fe] vs. [O/Fe],
[Al/Fe] vs. [Mg/Fe] and [Al/Fe] vs. [Na/Fe] for the sample stars:
indian red open stars, and Barbuy et al. (2007): orange filled squares.}
\label{anticorr} 
\end{figure*}

The odd-Z elements Na and Al show a different behaviour from each
other. Na shows no enhancements relative to the solar ratio,
and it is even deficient in one star, whereas Al is enhanced
in two stars. In the mean Na shows essentially a solar ratio with  
[Na/Fe]$\sim$-0.03, and Al is slightly enhanced with [Al/Fe]=+0.15. 
 Despite the small size of the sample, in Fig. \ref{anticorr}
we show [Na/Fe] vs. [O/Fe], [Al/Fe] vs. [O/Fe],
[Al/Fe] vs. [Mg/Fe] and [Al/Fe] vs. [Na/Fe], to check if
the expected anti-correlations Na-O, Al-O, Al-Mg, Al-Na are 
seen (Carretta et al. 2009; Renzini et al. 2015). In this figure
are also added the stars analysed in Barbuy et al. (2007).
Looking at the present results,
anti-correlations Al-O and Al-Mg might be present, but in the overall
 none seems clear. Let us add that NGC 6558 has an
absolute magnitude of M$_{\rm V}$ = -6.44 (Harris 1996), 
a little less massive than M4 (NGC 6121) with M$_{\rm V}$ = -7.19, the latter
clearly showing such anti-correlations. In conclusion, further data
are needed for NCG 6558, both spectroscopy for larger samples,
as well as UV photometry as carried out in Piotto et al. (2015),
in order to identify second generation stars in this cluster. 

The `bona-fide'  $\alpha$-elements O and Mg, produced in hydrostatic phases
of massive stars, with enhancements of [O/Fe]=+0.40 and [Mg/Fe]=+0.33, 
 indicate that they were enriched early in the Galaxy.
Si and Ca instead, that form mainly in explosive nucleosynthesis
in SN type II (Woosley et al. 1995, 2002), tend to be rather low, with
[Si/Fe]=+0.11, and [Ca/Fe]=+0.07.
 The same pattern was found for NGC~6522,
with [Si/Fe]=+0.13, and [Ca/Fe]=+0.13
(Barbuy et al. 2014), and HP~1, with 
[Si/Fe]=+0.27, and [Ca/Fe]=+0.13 (Barbuy et al. 2016). 
As for Ti that is in fact an iron-peak element, but behaves
very similarly to Si, Ca, the similarity with the other two clusters
 is also found:
[TiI/Fe]=+0.21, [TiII/Fe]=+0.24 in NGC~6558,
[TiI/Fe]=+0.16, [TiII/Fe]=+0.21 in NGC~6522, and
[TiI/Fe]=+0.04, [TiII/Fe]=+0.17 in HP~1.

This indicates that the moderately metal-poor clusters in the Galactic bulge
appear to show a typical abundance pattern. The elements Si, Ca, and Ti
 have lower abundances relative to O and Mg; this difference is more
pronounced than  in bulge field stars, where Si, Ca, and Ti 
tend to show abundances of about 0.1 dex lower than those of  O and Mg.
(e.g. Barbuy et al. 2018a; Bensby et al. 2017; Fria\c ca \& Barbuy 2017;  
McWilliam 2016).
The iron-peak elements were derived for these stars in Ernandes et al.
(2018).

As for the heavy elements, 
 we compared the present results with available literature
abundances of heavy elements in bulge field stars, and results for
NGC 6558 from Barbuy et al. (2007), and other globular clusters:
a) Johnson et al. (2012) derived abundances of  Eu
in common with the present element abundance derivation.
 Their observations concern
red giants in Plaut's field, located at l=-1$^\circ$, b=-8$^{\circ}$, and 
 l=-1$^\circ$, b=-$8\fdg5$, and their stars have metallicities
in the range $-$1.6 $\leq$ [Fe/H] $\leq$ +0.5;
b) Siqueira-Mello et al. (2016) analysed five stars indicated to have metallicities
around [Fe/H]$\sim$-0.8 in the ARGOS survey (Ness et al. 2013).
Abundances of the heavy elements
Y, Zr, La, Ba, and Eu were derived;
c) Bensby et al. (2017) present element abundances of 90 microlensed bulge
dwarfs and subgiants. Their study includes the 
abundances of the heavy elements Y and Ba;
d) van der Swaelmen et al. (2016) derived abundances of Ba, La, Ce, Nd and Eu
for 56 bulge field giants; 
e) five stars in NGC 6558 from Barbuy et al. (2007), where
abundances of Ba, La and Eu were derived;
f) As for other globular clusters, results for the similar clusters  
NGC~6522 (Barbuy et al. 2014), and HP~1 (Barbuy et al. 2016)
are included, as well as  M62 (NGC 6266), projected towards the bulge,
studied by 
Yong et al. (2014).

\begin{figure*}
\centering
\includegraphics[scale=0.8]{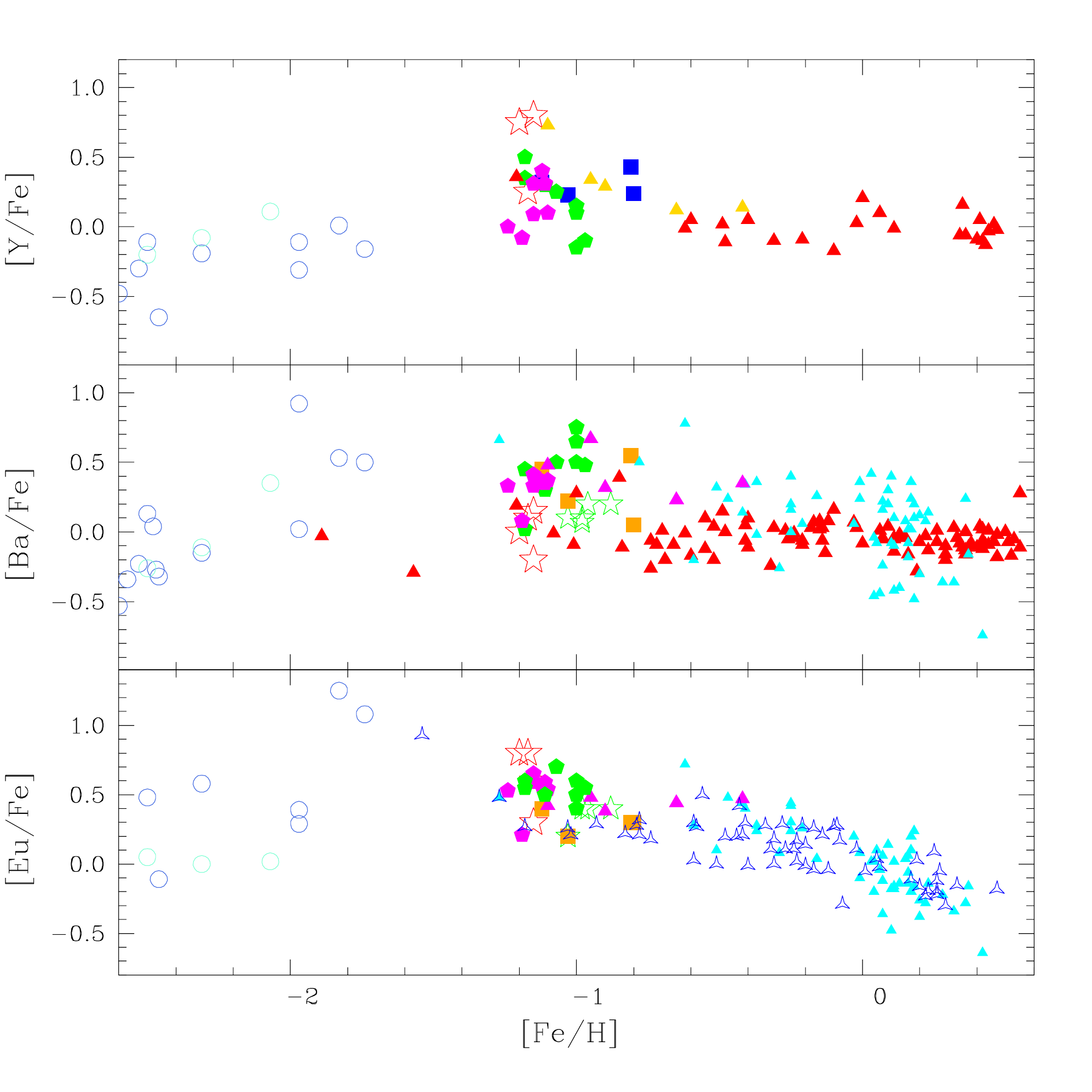}
\caption{[Y,Ba,Eu/Fe] vs. [Fe/H]. 
Symbols:
present results: red open stars; NGC 6558 (Barbuy et al.
2007): green open stars; NGC~6522 (Barbuy et al. 2014): blue filled squares;
HP~1 (Barbuy et al. 2016): green filled pentagons; 
dwarf microlensed stars (Bensby et al. 2017): red filled triangles;
field RGB stars (Siqueira-Mello et al. 2016): gold filled triangles;
field RGB stars (van der Swaelmen et al. 2016): cyan filled triangles;
field RGB stars (Johnson et al. 2012): blue open triangles;
M62 (Yong et al. 2014): magenta filled pentagons;
and metal-poor stars: (Howes et al. 2016): royal blue open circles;
(Casey \& Schlaufman 2015): sky blue open circles; 
(Koch et al. 2016): acqua-marine open circles.
}
\label{plotheavy} 
\end{figure*}

In Figure \ref{plotheavy} we plot the present results for the heavy elements 
Y, Ba, and Eu-over-Fe for the four  sample stars, compared with the above-cited references.
This figure indicates that overall Y and Ba show a spread
 at [Fe/H]$\sim$$-$1.0. 
 This spread in abundances is compatible with expectations 
from massive spinstars (see Chiappini et al. 2011, 
Chiappini 2013), based on predictions by Frischknecht et al. (2016).
 We note on the other hand that the Sr, Y abundances are derived mostly
from \ion{Sr}{I} and \ion{Y}{I} lines, whereas the dominant species 
are \ion{Sr}{II} and \ion{Y}{II},
bringing a non-negligible uncertainty in these values.

The second-peak s-element Ba is low, with a solar ratio,
confirming earlier solar-like ratios of Ba-to-Fe and La-to-Fe 
from Barbuy et al. (2007). Instead, 
the first-peak elements Y and Sr appear enhanced, from the few lines
measurable. This latter result has to be taken with caution, given
the noise in the spectra, and the use of neutral species, as mentioned
above. If confirmed, this would show
 once more the difference
in nucleosynthesis processes of the first peak elements, that
often show enhancements, in a non-identified process so-called
'lighter element primary process (LEPP)' 
(mainly Sr, Y, and Zr) process (Travaglio et al. 2004, Bisterzo et al. 2017).
The excess of the first-peak heavy-elements could be due to
neutrino-driven winds arising in SNII
(Woosley et al. 1994), electron-capture supernovae
(Wanajo et al. 2011), among other possibilities.

The r-element Eu is enhanced with a mean [Eu/Fe]$\sim$+0.60, compatible 
with production by supernovae type II. Presently the most
suitable source of r-elements in SNII, is
through neutron star mergers, with nucleosynthesis
occurring during the merging and milliseconds afterwards 
(e.g. Goriely et al. 2015, Wanajo et al. 2014).
Pian et al. (2017) has confirmed the presence of
 {\it r}-process elements in the neutron star merger
GW170817 revealed by LIGO - 
Laser Interferometer Gravitational-Wave Observatory\footnote{http://www.ligo.caltech.edu}  and VIRGO\footnote{http://www.virgo-gw.eu} experiments.

 We note that high Eu-enhancements in
 globular clusters are not usual. The case of the
Eu-enhanced NGC~5986 (Johnson et al. 2017) might be similar to NGC~6558
in this respect. Enhancements of [Eu/Fe]$\sim$+0.4 are usually found
in field halo and globular cluster stars.
As pointed out by Johnson et al. (2017), higher enhancements are instead
 more typical of dwarf galaxies, such as the so-called r-process
ultra-faint dwarf galaxy Reticulum II (Ji et al. 2016), 
where [Eu/Fe]$\simgreat$+1.7 was derived.
The other possible similarity with NGC~5986,  is a possible La-enhancement,
however it is not clear that this is the case for NGC~6558, and should
be further investigated. Moreover, a high La abundance may be incompatible
with a low Ba abundance, noting that Ba abundances were not derived in
NGC~5986. 

\begin{figure*}
\centering
\includegraphics[scale=0.8]{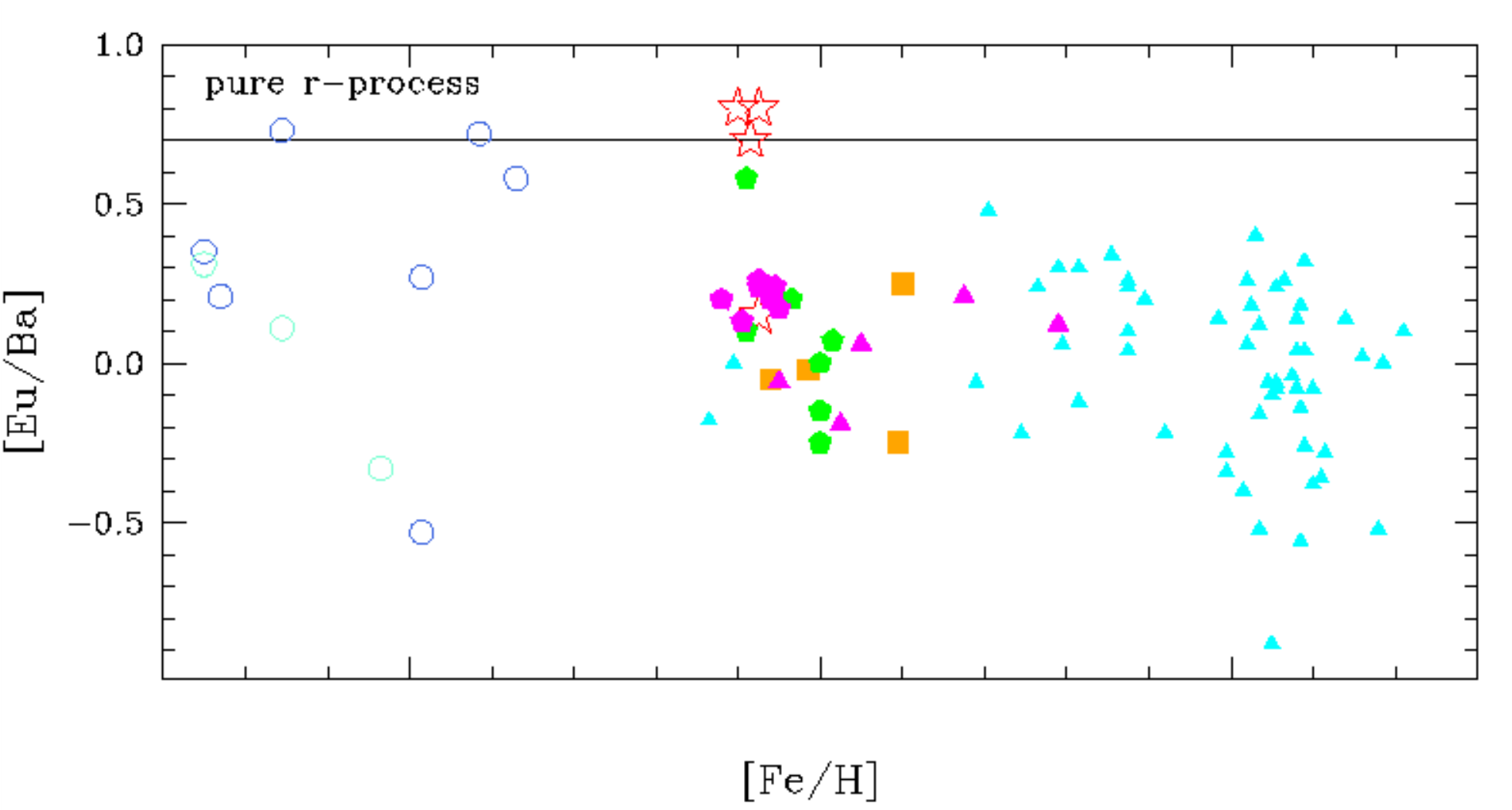}
\caption{ [Eu/Ba] vs. [Fe/H]. Same  symbols as in Fig. \ref{plotheavy}.
}
\label{baeu} 
\end{figure*}

The Eu/Ba ratio is indicative of r- to s-process, where a ratio
[Eu/Ba]$\sim$0.7 points to a nucleosynthetic production through the main
 r-process (e.g. Battistini \& Bensby 2016, Trevisan \& Barbuy 2014).
In Fig. \ref{baeu}we show [Eu/Ba] for the same stars as in
Fig. \ref{plotheavy} for which both Ba and Eu abundances are available.
It is interesting to see that  three among the four sample stars
show ratios compatible with the r-process, and a
 few other more metal-poor stars are also located in the pure r-process
case, whereas most of the stars in the figure are not,
their heavy-element abundances being
more compatible with a mixture of r- and s-processes.

\section{Conclusions}

We present photometric V and I data, and high resolution
spectroscopic data for four stars in NGC 6558.
With a set of 
subarcsec ($\sim$0.5'') seeing images obtained at the NTT, 
 with a time difference of 19 years, proper motion
decontamination, with high accuracy, was applied.
These photometric  data have allowed us to
derive reddening and distance, and provided
 the basis to derive photometric stellar parameters. 
Based on the proper motions obtained from the combination of
these NTT images, by Rossi et al. (2015), the orbital analysis by
 P\'erez-Villegas et al. (2018) has shown that NGC 6558 is trapped by the bar,
 having an orbit with a radial extension between 0.1 and 2.5 kpc and a
 maximum vertical excursion of z$\sim$1.4 kpc. 
These orbital parameters are similar to those of the globular cluster 
HP 1, and approximately two times larger than 
NGC 6522 (P\'erez-Villegas et al. 2018). 
NGC 6522 and HP 1 have similar ages of ~13 Gyr 
(Kerber et al. 2018a,b), pointing out that these
clusters could be among the 
oldest objects in the Galaxy, trapped at some point in the bar/bulge.

\begin{figure*}
\centering
\includegraphics[scale=0.8]{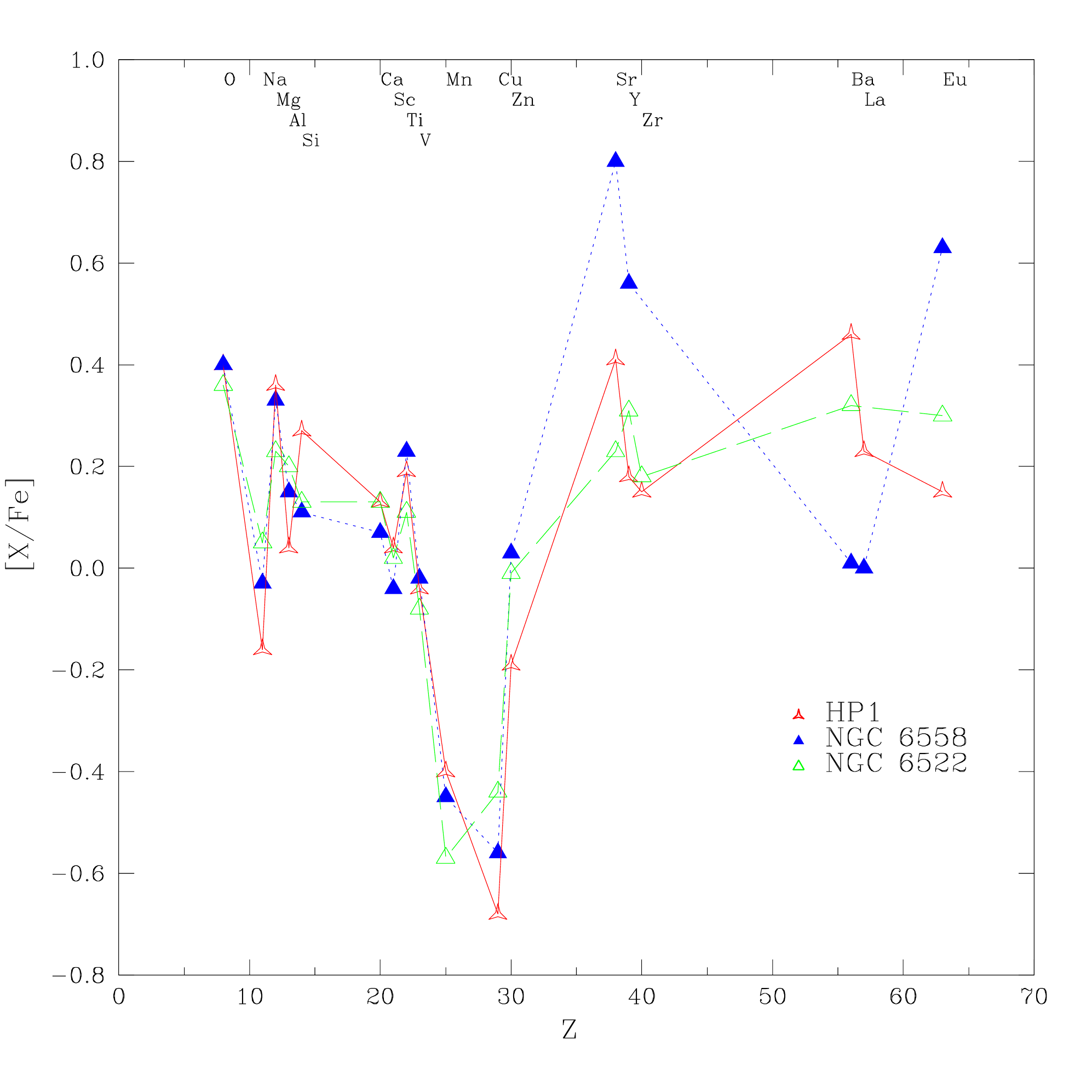}
\caption{Abundance pattern [X/Fe] vs. atomic number Z of
the moderately metal-poor bulge globular clusters NGC~6558,
NGC~6522 and HP~1. Symbols: blue filled triangles: NGC~6558;
green open triangles: NGC~6522;
red open triangles: HP~1.
}
\label{pattern} 
\end{figure*}

The UVES spectroscopic data allowed to derive the final
stellar parameters and abundance ratios for the four sample stars.
We note that the abundance pattern of NGC 6558 
is very similar to those of NGC 6522
(Barbuy et al. 2014) and HP 1 (Barbuy et al. 2016), in many aspects, 
as illustrated in Figure \ref{pattern}. The iron-peak elements
in these clusters are also included in the comparison,
as derived in Ernandes et al. (2018).
The similarities found in the three clusters are:
relatively low odd-Z elements Na and Al;
normally enhanced $\alpha$-elements O, Mg; low enhancements of Si, Ca;
moderate enhancements of the iron-peak ($\alpha$-like) Ti;
low abundances of the iron-peak elements Mn, Cu, and Zn, where Zn in particular
differs from values in bulge field stars (e.g. da Silveira et al. 2018,
Barbuy et al. 2015, 2018); relatively high first-peak of heavy elements;
variable second-peak heavy elements; variable enhancements of the r-element
Eu.

With the present work we are able to gather the analysis of the three 
main representatives of
moderately metal-poor and BHB globular clusters in the bulge,
and to compare their shared properties. Further photometric
and spectroscopic studies of the
oldest globular clusters in the bulge would be of great interest.

\begin{acknowledgements}
BB and EB acknowledge acknowledge grants from CNPq, 
CAPES - Finance code 001, and Fapesp. 
LM, HE acknowledge CNPq/PIBIC fellowships.
LK acknowledges a CNPq postdoctoral fellowship and the
 Fapesp grant 2016/24814-5. APV acknowledges the 
Fapesp postdoctoral fellowship no. 2017/15893-1.
We thank Elvis Cantelli for data reduction.
SO acknowledges the Dipartimento di Fisica e Astronomia dell'Universit\`a di
Padova.
\end{acknowledgements}


\end{document}